\documentclass[12pt]{article}

\usepackage{newtxtext,newtxmath}

\usepackage{graphicx}

\usepackage[letterpaper,margin=1in]{geometry}

\renewenvironment{abstract}
	{\quotation}
	{\endquotation}

\date{}

\makeatletter
\renewcommand{\fnum@figure}{\textbf{Figure \thefigure}}
\renewcommand{\fnum@table}{\textbf{Table \thetable}}
\makeatother

\usepackage{scicite}

\usepackage{url}

\newcommand{\diff}{\mathrm{d}}
\renewcommand{\vec}[1]{\mathbf{#1}}
\newcommand{\cell}[1]{\begin{array}{c}#1\end{array}}

\def\scititle{
	Resolving coupled transport in space and time from molecular fluctuations in confined fluids
}
\title{\bfseries \boldmath \scititle}

\author{
	Minh-Thê~Hoang~Ngoc$^{1\ast}$,
	Ian C. Bourg$^{1,2}$\and
	\small$^{1}$Department of Civil and Environmental Engineering, Princeton University, Princeton, NJ 08540, USA.\and
	\small$^{2}$High Meadows Environmental Institute, Princeton University, Princeton, NJ 08540, USA.\and
	\small$^\ast$Corresponding author. Email: mh7003@princeton.edu\and
}

\begin{document} 

\maketitle

\begin{abstract} \bfseries \boldmath
Transport in fluids is generally reduced to continuum laws parametrized by bulk coefficients and effective interfacial parameters, such as viscosities, diffusivities, slip lengths, and interfacial resistances. This description becomes incomplete at the nanoscale, where spatial heterogeneity, molecular structure, and finite relaxation times are inseparable from the transport process. Here we formulate coupled transport in nanoconfined fluids as a space--time-resolved Onsager response matrix and extract it from equilibrium molecular dynamics simulations. Applied to a confined charged fluid, the framework resolves the nonlocal and transient pathways coupling particle, solute, heat, and charge transport. Momentum transport appears as a long-lived, nonlocal hydrodynamic mode, whereas charge transport relaxes rapidly through localized ionic friction. Off-diagonal responses reveal distinct projected dynamics, providing a microscopic basis for nonlocal, history-dependent transport laws.
\end{abstract}


\newpage

\noindent
Most of our intuition about transport is built around a small set of coefficients that linearly relate fluxes to driving forces~\cite{bird2007transport}. In electromagnetism, however, this picture has long been superseded: material response is described as inherently nonlocal and dynamical, through wavevector- and frequency-dependent functions such as the dielectric tensor $\boldsymbol{\varepsilon}_r(\vec{k},\omega)$~\cite{jackson1999classical}. Heterogeneity, memory, and dispersion are not corrections, but central features of the description. By contrast, fluid transport—outside of specialized areas such as rheology~\cite{larson1999structure}—is still most often described using homogeneous and stationary coefficients, even when spatial structure and temporal relaxation are dominant~\cite{hoang2023frequency}.

This limitation becomes acute in nanoconfined fluids, where interfacial structure, molecular heterogeneity, and finite relaxation times predominate, with first-order impacts across geoscience, biology, and energy technologies~\cite{bocquet2010nanofluidics,simon2020perspectives,Kleber2021dynamics,wild2022reverse}. Under molecular-scale confinement, transport cannot be cleanly separated into bulk behavior and boundary contributions: interfacial dynamics and fluid transport are intrinsically coupled, challenging both the interpretation of effective parameters and the formulation of boundary conditions~\cite{kavokine2021fluids,schlaich2025theory,hartkamp2018surfacecharge}.

A number of concrete challenges illustrate this breakdown. Near charged surfaces, the relaxation of ions in the electrical double layer (EDL) involves coupled electrostatic, hydrodynamic, and thermal fluctuations that are difficult to capture within static or local descriptions~\cite{marbach2018transport,kavokine2022fluctuations}. Similarly, coupled transport processes central to electrokinetics and thermo-osmosis are typically described through Onsager matrices, yet their determination from molecular simulations remains indirect and often ambiguous~\cite{yoshida2014generic,marbach2019osmosis,helms2023intrinsic,hoang2026coupled}. More generally, the physical meaning of effective parameters such as slip lengths, zeta potentials, or interfacial resistances becomes unclear when interfacial and bulk dynamics are strongly coupled.

These observations point to a central question: how does transport emerge from molecular fluctuations in space and time under confinement? While statistical mechanics formally provides this connection through space- and time-dependent correlation functions, in practice transport is almost always reduced to integrated quantities, yielding effective coefficients or asymptotic descriptions~\cite{zwanzig2001nonequilibrium,tevrugt2020DDFT}. As a result, the spatiotemporal structure of transport—where heterogeneity, memory, and interfacial effects originate—remains largely inaccessible.

Recent advances in simulations and experiments are beginning to lift this limitation~\cite{franosch2011resonances,mangaud2020sampling,pireddu2024impedance,Underwood2022dielectric,domingues_diffusivity_tensors_2024,hoang2023ionicfluctuations,mackay2024countoscope}. This opens the possibility of a different perspective: rather than reducing transport to coefficients, one could resolve how it emerges across space and time from microscopic fluctuations.

Here we show that coupled transport in nanoconfined fluids can be formulated as a space--time-resolved generalization of the Onsager response matrix. We test this idea using equilibrium molecular dynamics (MD) simulations of a Lennard--Jones fluid confined between charged walls, where local particle, solute, heat, and charge fluxes can be sampled directly. In this framework, transport emerges from spatially heterogeneous, time-dependent correlations, and conventional transport coefficients appear as fully integrated limits of the corresponding space--time response kernels. This provides a transparent physical picture of how transport develops across scales.

\subsection*{From transport coefficients to space--time response kernels}\label{subsec:kernel}
Macroscopic transport is usually described through a finite set of coefficients relating integrated fluxes \(J_a\) to driving forces \(F_b=-\nabla\phi_b\), where \(\phi_b\) is the thermodynamic potential conjugate to the transported quantity \(b\). These coefficients are assembled into the Onsager matrix \(L_{ab}\), whose diagonal entries describe direct transport processes and whose off-diagonal entries describe coupled responses such as electro-, thermo-, and diffusio-osmosis~\cite{onsager_recip_1931}:
\begin{equation}
    J_a = - \sum_b L_{ab}\,\nabla \phi_b .
    \label{eq:Onsager_matrix}
\end{equation}
The relation between interfacial structures and the \(L_{ab}\) coefficients is a key motivation for MD simulations~\cite{maduar2015electrohydrodynamics,ganti2017thermoosmosis}. While this description is often sufficient at macroscopic scales, it averages over the spatial heterogeneity and temporal relaxation that become central under nanoconfinement. We therefore focus on the object that precedes this coarse graining: the space--time response kernel.

For spatially varying and time-dependent perturbations around a stationary equilibrium state, the Onsager flux--force relation can be promoted to a linear constitutive relation for local flux densities that incorporates memory and nonlocal effects:
\begin{equation}
    \left\langle \vec{j}_a(\vec{r},t)\right\rangle_\mathrm{neq}
    =
    - \sum_b \int_V \int_{-\infty}^t
    \vec{R}_{ab}(\vec{r},\vec{r}',t-t')
    \cdot \vec{\nabla} \phi_b(\vec{r}',t')
    \, \mathrm{d}t'\,\mathrm{d}\vec{r}' .
    \label{eq:LRT_field}
\end{equation}
Here \(\vec{j}_a(\vec r,t)\) is the instantaneous local flux density whose spatial integral gives the macroscopic flux $J_a$ up to the geometrical normalization appropriate to the chosen flux--gradient convention. The kernel \(\vec{R}_{ab}(\vec r,\vec r',t)\) is therefore the space--time resolved counterpart of \(L_{ab}\): it measures the local response of the flux of quantity \(a\) at \((\vec r,t)\) to the local gradient of the potential conjugate to quantity \(b\) at \((\vec r',t')\). For homogeneous and stationary perturbations, the coefficient-level description is recovered by integrating the kernel over space and time:
\begin{equation}
    \vec{L}_{ab}
    =
    \lim_{t\to\infty}
    \int_0^t
    \int_V
    \int_V
    \vec{R}_{ab}(\vec{r},\vec{r}',s)
    \, \mathrm{d}\vec{r}\,
    \mathrm{d}\vec{r}'\,
    \mathrm{d}s .
    \label{eq:GK_KB}
\end{equation}
More generally, different degrees of spatial and temporal coarse graining define a hierarchy of response functions.
\begin{small}
\[
\begin{array}{c||c|c|c}
  & \text{Differential}
  & \text{Integrated}
  & \text{Stationary} \\
\hline
\hline
\text{Local--Local}
  & \cell{\vec{R}_{ab}(\vec{r},\vec{r}',t)}
  & \cell{
  \vec{\mathcal{G}}_{ab}(\vec{r},\vec{r}',t) \\
  = \displaystyle\int_0^t
  \vec{R}_{ab}(\vec{r},\vec{r}',s)\,\mathrm{d}s}
  & \cell{
  \vec{G}_{ab}(\vec{r},\vec{r}') \\
  = \displaystyle\lim_{t\to\infty}
  \vec{\mathcal{G}}_{ab}(\vec{r},\vec{r}',t)}
  \\[6pt]
\hline
\text{Local--Global}
  & \cell{
  \vec{K}_{ab}(\vec{r},t) \\
  = \displaystyle\int_V
  \vec{R}_{ab}(\vec{r},\vec{r}',t)\,\mathrm{d}\vec{r}'}
  & \cell{
  \vec{\mathcal{M}}_{ab}(\vec{r},t) \\
  = \displaystyle\int_0^t
  \vec{K}_{ab}(\vec{r},s)\,\mathrm{d}s}
  & \cell{
  \vec{M}_{ab}(\vec{r}) \\
  = \displaystyle\lim_{t\to\infty}
  \vec{\mathcal{M}}_{ab}(\vec{r},t)}
  \\[6pt]
\hline
\text{Global--Global}
  & \cell{
  \vec{C}_{ab}(t) \\
  = \displaystyle\int_V
  \vec{K}_{ab}(\vec{r},t)\,\mathrm{d}\vec{r}}
  & \cell{
  \vec{\mathcal{L}}_{ab}(t) \\
  = \displaystyle\int_0^t
  \vec{C}_{ab}(s)\,\mathrm{d}s}
  & \cell{
  \vec{L}_{ab} \\
  = \displaystyle\lim_{t\to\infty}
  \vec{\mathcal{L}}_{ab}(t)}
\end{array}
\]
\end{small}

Because the kernel couples two positions, a time delay, and two transported quantities, directly measuring \(\vec{R}_{ab}(\vec{r},\vec{r}',t)\) as a response function would be extremely arduous. Linear response theory~\cite{methods} provides a way around this difficulty by relating the space--time response kernel to equilibrium current correlations:
\begin{equation}
    \vec{R}_{ab}(\vec{r},\vec{r}',t-t')
    =
    \frac{1}{k_B T}
    \left\langle
    \vec{j}_a(\vec{r},t)
    \,
    \vec{j}_b^\dagger(\vec{r}',t')
    \right\rangle_\mathrm{eq}.
    \label{eq:LRT_FDT_field}
\end{equation}
Together with the coarse-graining relation in Eq.~\eqref{eq:GK_KB}, Eq.~\eqref{eq:LRT_FDT_field} shows that Green--Kubo and Kirkwood--Buff-type constructions~\cite{TOSL,zwanzig2001nonequilibrium} correspond to complementary projections of field-level equilibrium correlations. In this sense, these familiar quantities are not separate from the present framework, but correspond to particular projections of the same underlying space--time correlation structure.

The field-level formulation also exposes fundamental symmetry relations that remain after full coarse graining. In stationary systems with microscopic time-reversal symmetry, the response kernel satisfies~\cite{kubo1957}
\begin{equation}
    \vec{R}_{ab}(\vec r,\vec r',t)
    =
    \vec{R}_{ba}^{\dagger}(\vec r',\vec r,-t)=\vec{R}_{ba}^{\dagger}(\vec r',\vec r,t),
    \label{eq:time_onsager_kernel}
\end{equation}
so that the fully integrated coefficients recover the Onsager reciprocal symmetry
\begin{equation}
    \vec{L}_{ab}
    =
    \vec{L}_{ba}^{\dagger}.
    \label{eq:onsager_integrated_limit}
\end{equation}

Here, we use the same relation operationally, but stop before the usual spatial averaging, time integration, and long-time limit. The pre-integrated current correlations are not treated as intermediate quantities on the way to \(L_{ab}\); they are the central observables of the paper.

We implement this framework in a minimal model of a charged nanopore (Fig.~\ref{fig:diag_responses}A). The system consists of a partially charged Lennard--Jones (LJ) fluid confined between two negatively charged walls separated by a distance \(H=20\sigma\). The simulation cell has lateral dimensions \(50\sigma \times 50\sigma\), and contains approximately \(3.1\times10^4\) particles at temperature \(T=1.4\) and total density \(\rho_{\mathrm{tot}}\simeq 0.64\). The charged species are monovalent, \(q_\pm=\pm e\), with excess counterions added to ensure global electroneutrality, and a charged-particle fraction \(\rho_s/\rho_{\mathrm{tot}}\simeq 0.25\). The walls carry a negative surface charge density \(\Sigma=-0.1\,e/\sigma^2\). Unless otherwise stated, all quantities are reported in reduced LJ units. All plotted quantities are rescaled for visualization to emphasize their spatio-temporal structure. Their amplitudes are constrained by the Onsager coefficients reported in Tab.~\ref{sup_tab:L_ab_numerical}, obtained through the successive spatio-temporal integrations (Eq.~\eqref{eq:GK_KB}). Further details are provided in Materials and Methods~\cite{methods}.

\subsection*{Hydrodynamic and charge transport follow distinct response pathways}
\label{subsec:diag}
We first illustrate the extracted response hierarchy using the two diagonal sectors \(nn\) and \(cc\), corresponding respectively to hydrodynamic fluid and ionic charge transport. This comparison provides the simplest physical demonstration of the framework: the equilibrium fluctuation analysis captures both long-lived momentum transport across the pore and faster, more localized charge relaxation in the EDL.

\begin{figure}
    \centering
    \begin{minipage}[t]{.53\textwidth}
        \centering
        {\Large\bfseries A}\\[0.1em]
        \includegraphics[width=\linewidth]{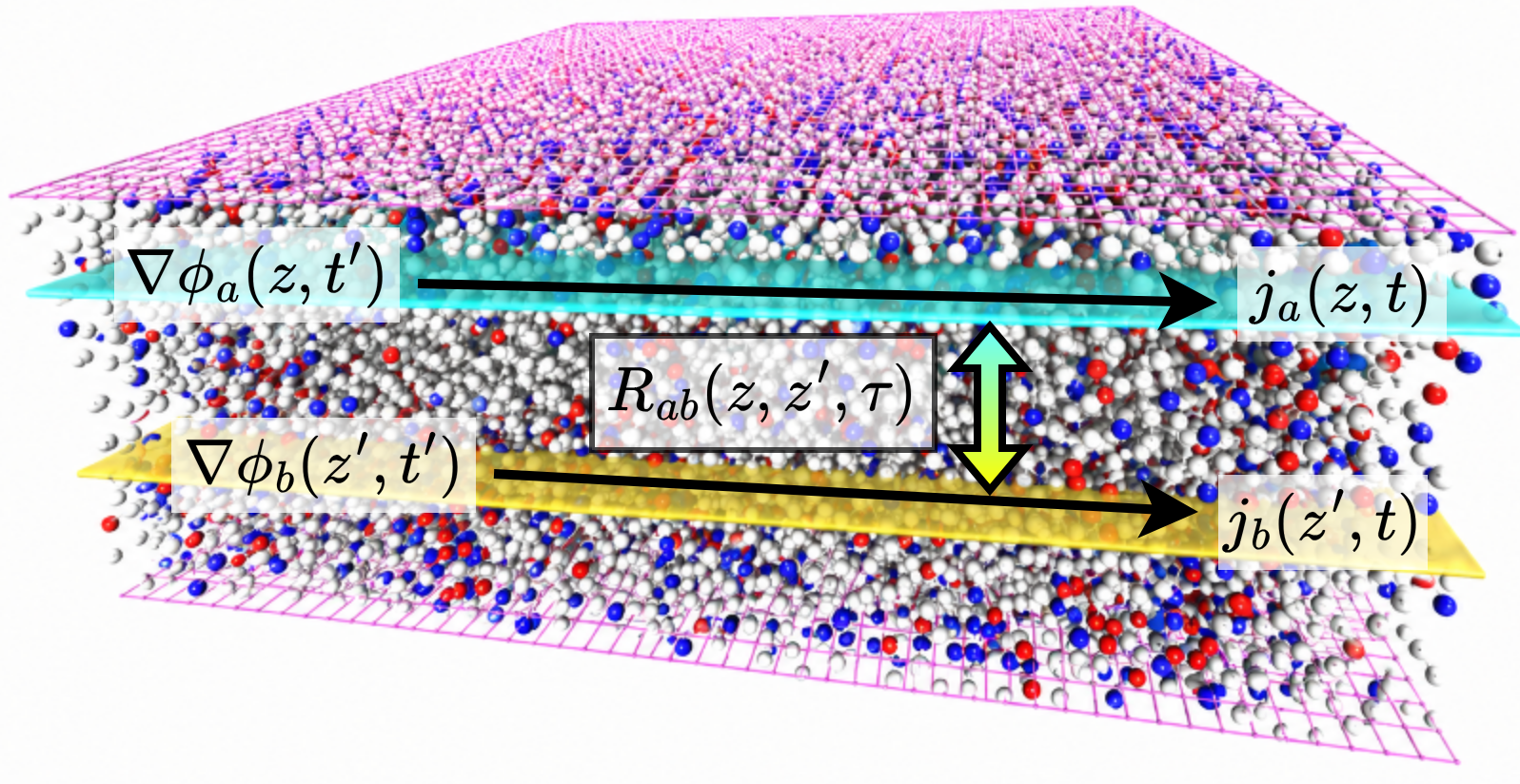}
    \end{minipage}
    \hfill
    \begin{minipage}[t]{.45\textwidth}
        \centering
        {\Large\bfseries B}\\[0.1em]
        \includegraphics[width=\linewidth]{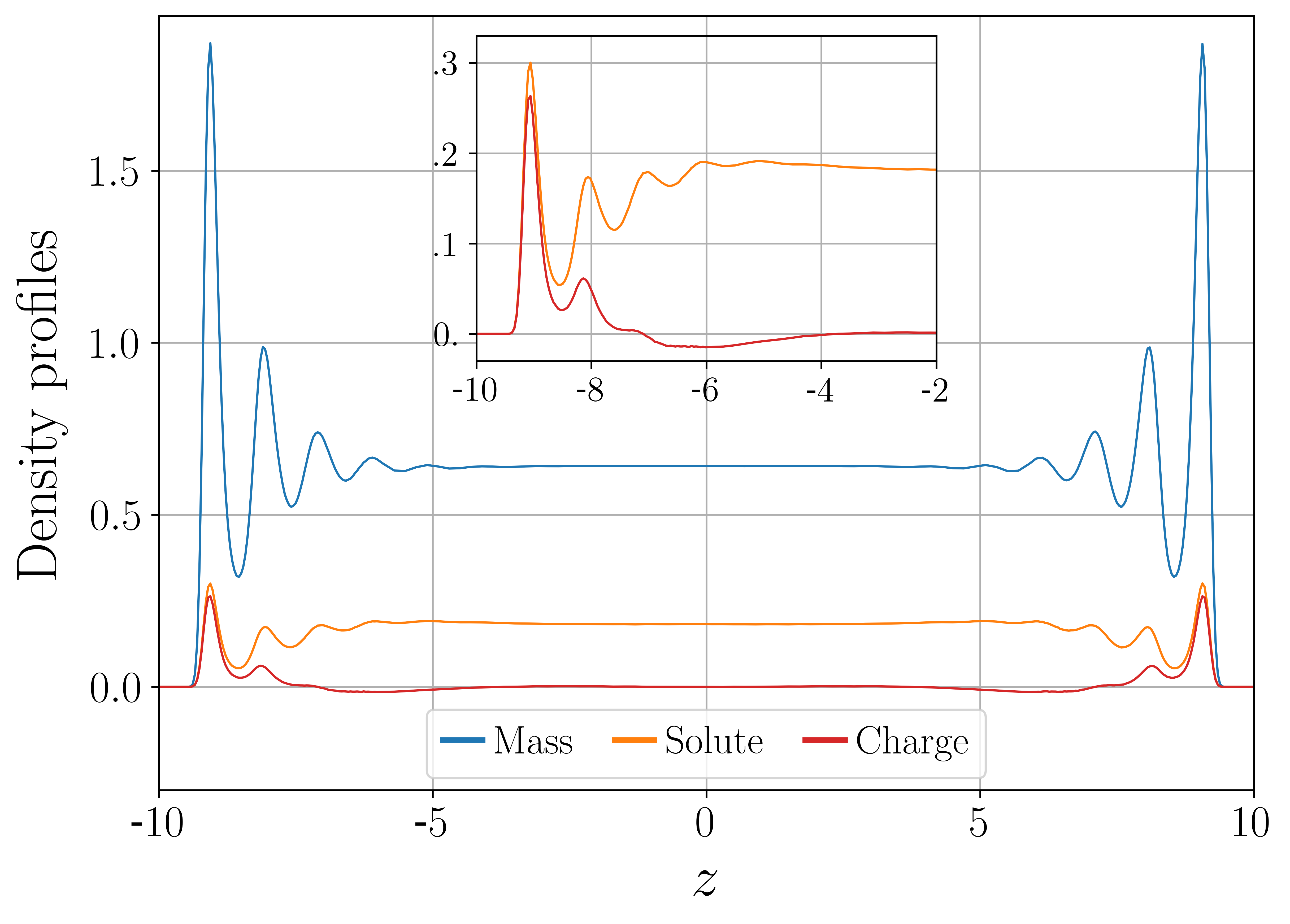}
    \end{minipage}
    
    \vspace{0.3em}

    \begin{minipage}[t]{.49\textwidth}
        \centering
        {\Large\bfseries C}\\[-0.em]
        \includegraphics[width=\linewidth]{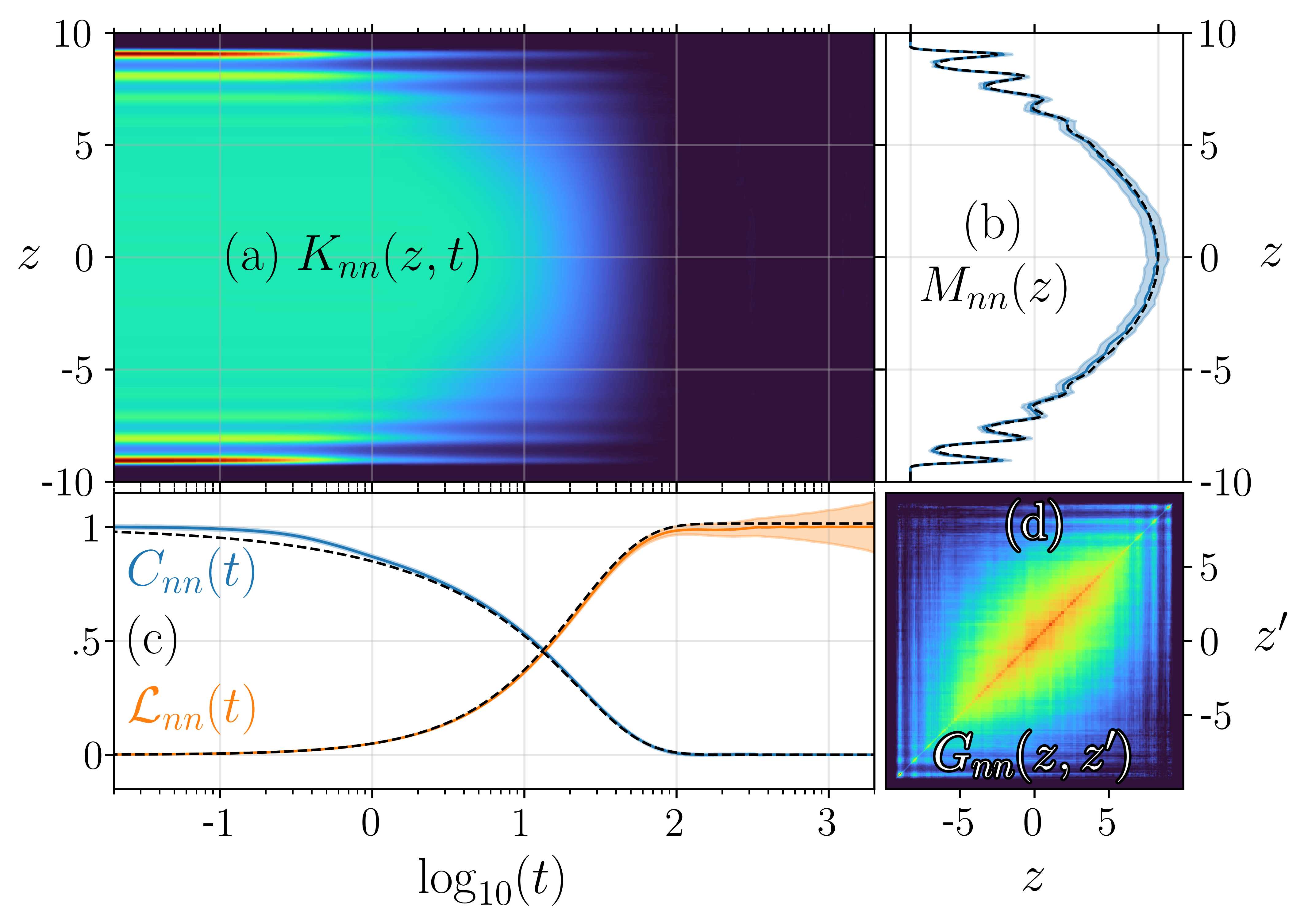}
    \end{minipage}
    \hfill
    \begin{minipage}[t]{.49\textwidth}
        \centering
        {\Large\bfseries D}\\[-0.em]
        \includegraphics[width=\linewidth]{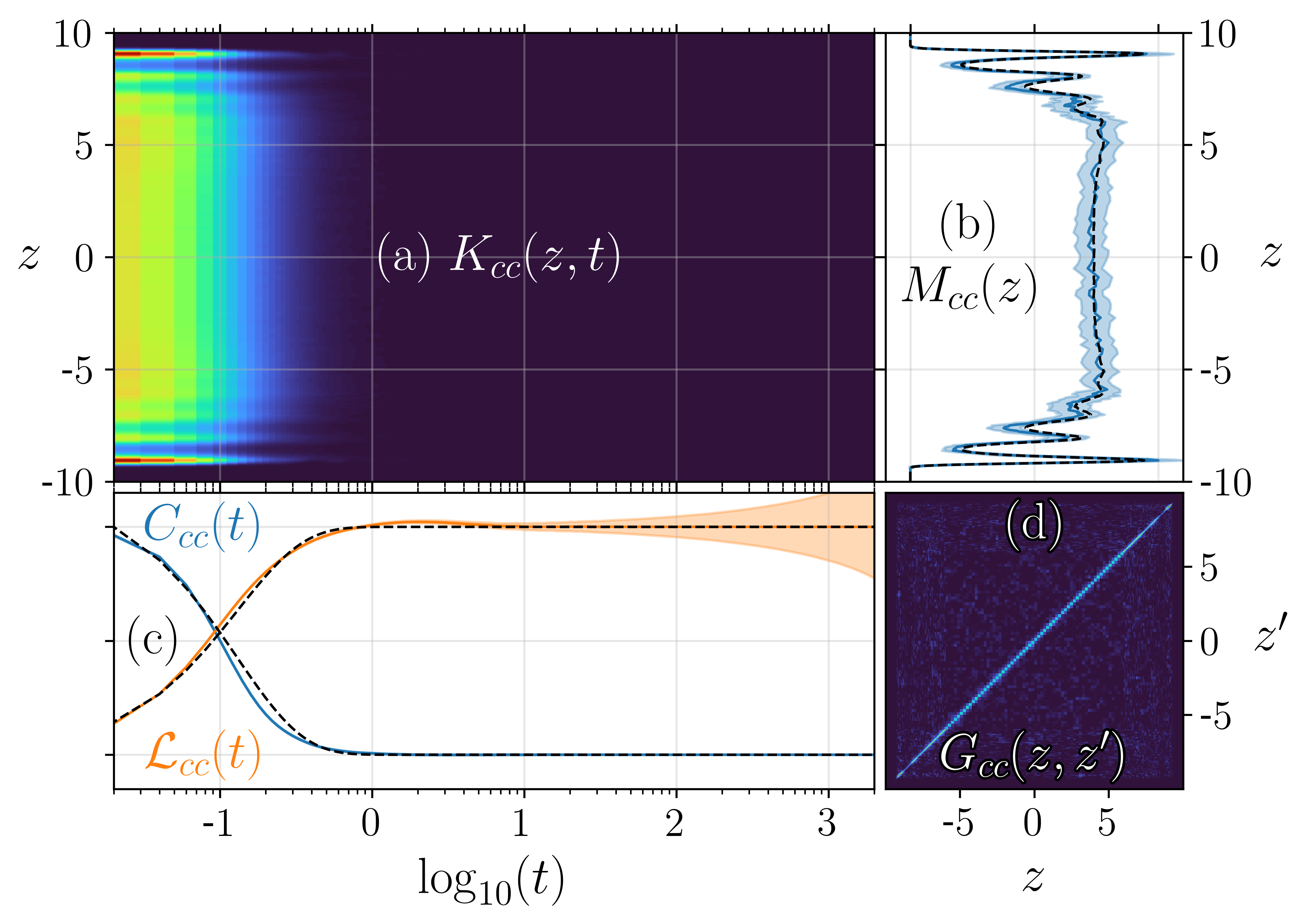}
    \end{minipage}
    \caption{\begin{small}
    \textbf{Hydrodynamic and charge transport emerging from molecular correlations.}
    (\textbf{A}) Snapshot of the charged LJ fluid confined between two charged walls. Transport is measured parallel to the walls and response functions are resolved along the confinement direction \(z\).
    (\textbf{B}) Equilibrium profiles of mass, solute and charge density, showing molecular layering near the walls. Inset shows interfacial charge organization with oscillations, commonly observed in systems with significant ion correlation effects. Uncertainties are smaller than the line thickness.
    (\textbf{C}) Hydrodynamic response hierarchy from the particle-flux correlation. (a) The main colormap shows the local--global kernel \(K_{nn}(z,t)\), with \(z=\pm10\) marking the wall positions and \(t\) shown on a logarithmic scale. The colormap is normalized to the maximum value, with dark blue indicating zero response and warmer colors indicating larger response. (b) Stationary profile \(M_{nn}(z)\). (c) Global correlation \(C_{nn}(t)\) and its cumulative integral \(\mathcal{L}_{nn}(t)\). (d) Local--local response \(G_{nn}(z,z')\). Shaded regions indicate statistical uncertainty; dashed black lines show analytical hydrodynamic predictions (see Materials and Methods).
    (\textbf{D}) Charge response hierarchy from the charge-flux correlation, shown with the same projections and color convention as in panel C.
    \end{small}}
    \label{fig:diag_responses}
\end{figure}

The fluid response \(K_{nn}(z,t)\) (Fig.~\ref{fig:diag_responses}C(a)) shows how hydrodynamic transport emerges from spatiotemporal molecular fluctuations. At short times, it exhibits pronounced interfacial features that follow the fluid density layering. These near-wall contributions relax rapidly, whereas the pore center exhibits a broader, long-lived response associated with viscous momentum diffusion. After spatial integration, this slow relaxation appears in the global correlation \(C_{nn}(t)\) and in the gradual convergence of its cumulative integral \(\mathcal{L}_{nn}(t)\) (Fig.~\ref{fig:diag_responses}C(c)).

Conversely, time integration of \(K_{nn}(z,t)\) yields the stationary response profile \(M_{nn}(z)\) (Fig.~\ref{fig:diag_responses}C(b)). This profile approaches a continuum-like hydrodynamic form in the central region, while retaining molecular-scale structure close to the walls. In this sense, the hydrodynamic response is not imposed as a continuum assumption: it emerges as the long-time limit of equilibrium molecular correlations. This provides a direct microscopic definition of the fluid response near the interface, reducing the ambiguity associated with continuum boundary conditions and effective wall positions~\cite{shi2025incorporating}. Dashed lines are obtained by fitting the transient approach to this profile using an unsteady no-slip Stokes description. This allows the effective confinement and viscosity to be estimated with substantially greater precision than standard approaches based only on the stationary mobility profile (see Materials and Methods~\cite{methods} and Fig.~\ref{sup_fig:fit_stokes}).

The charge response kernel \(K_{cc}(z,t)\) (Fig.~\ref{fig:diag_responses}D(a)) displays a qualitatively different organization. At very short times, it also reflects the interfacial layering of the confined electrolyte. However, unlike the hydrodynamic response, these correlations decay rapidly and remain spatially localized. This faster relaxation is visible in the global correlation \(C_{cc}(t)\) and in the rapid convergence of \(\mathcal{L}_{cc}(t)\) (Fig.~\ref{fig:diag_responses}D(c)). The stationary profile \(M_{cc}(z)\) (Fig.~\ref{fig:diag_responses}D(b)) retains interfacial layering but becomes nearly flat in the pore center, consistent with a bulk-like ionic conductivity in the homogeneous region.

This contrast reflects different dissipation mechanisms. Momentum transport proceeds through viscous diffusion across the pore and is ultimately dissipated by wall friction. Charge transport is instead governed by ionic friction and electrostatic relaxation from local charge separation. Time integration therefore yields an ionic conductivity profile combining interfacial structuring with an approximately homogeneous bulk contribution. This behavior is qualitatively consistent with a local Smoluchowski--Nernst--Einstein picture, while the full response kernel retains the correlated ionic motions that can make Nernst--Einstein descriptions incomplete~\cite{bazant2004diffuse,fong2021ioncorrelations}.

The local--local responses \(G_{nn}(z,z')\) and \(G_{cc}(z,z')\), shown in Fig.~\ref{fig:diag_responses}C(d) and D(d), provide a complementary view of these mechanisms. They quantify the spatial nonlocality of dissipation and response. The hydrodynamic sector displays extended correlations across the pore, consistent with nonlocal momentum diffusion, whereas the charge sector is much more localized, with correlations concentrated close to the diagonal \(z=z'\).

\subsection*{Electrokinetic coupling emerges from nonlocal interfacial correlations}\label{subsec:eof_sc}
\begin{figure}
    \centering
    \begin{minipage}[t]{.45\textwidth}
        \centering
        {\Large\bfseries A}\\[0.1em]
        \includegraphics[width=\linewidth]{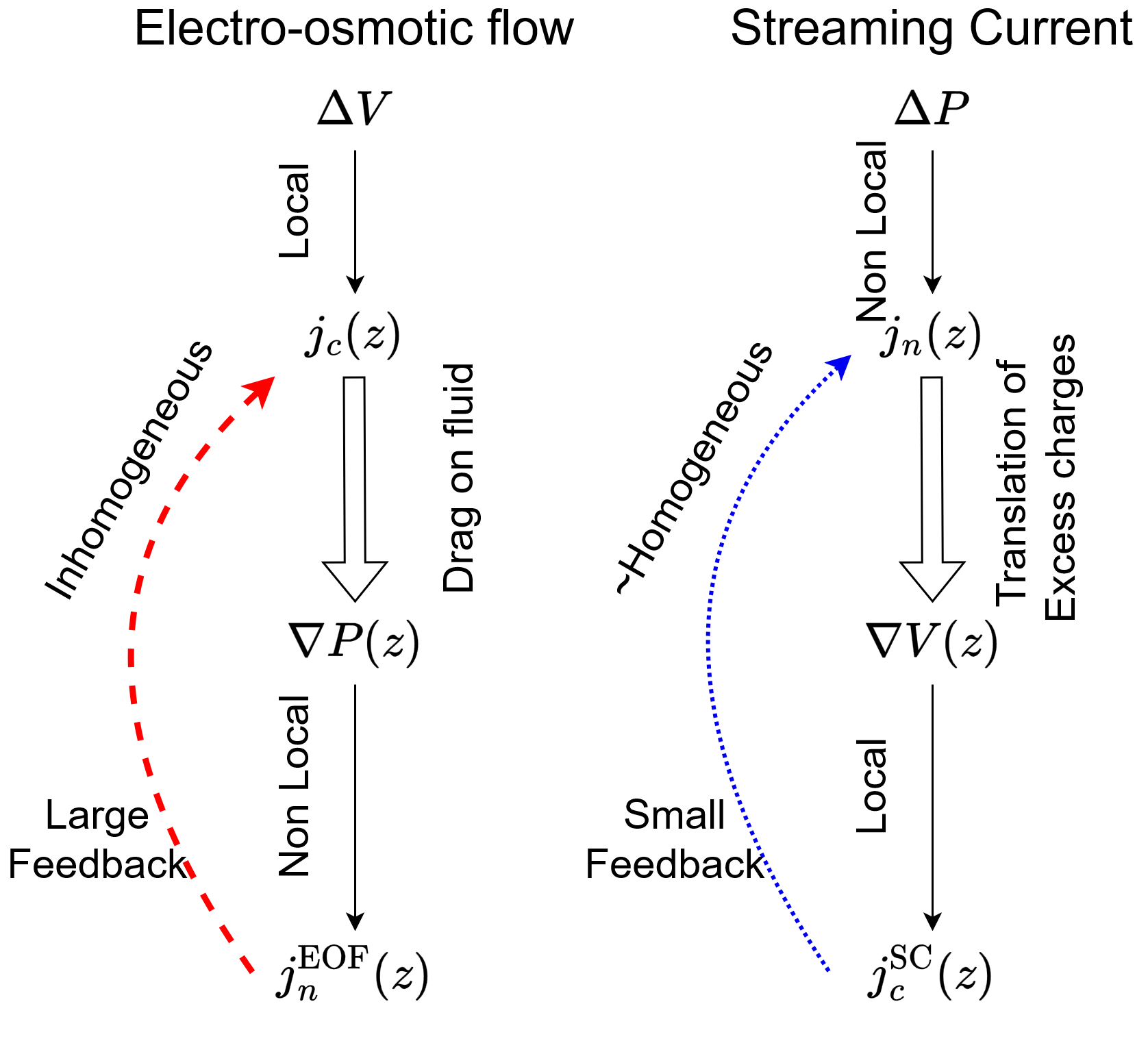}
    \end{minipage}
    \hspace{0.01\textwidth}
    \begin{minipage}[t]{0.52\textwidth}
        \centering
        {\Large\bfseries B}\\[0.1em]
        \includegraphics[width=\linewidth]{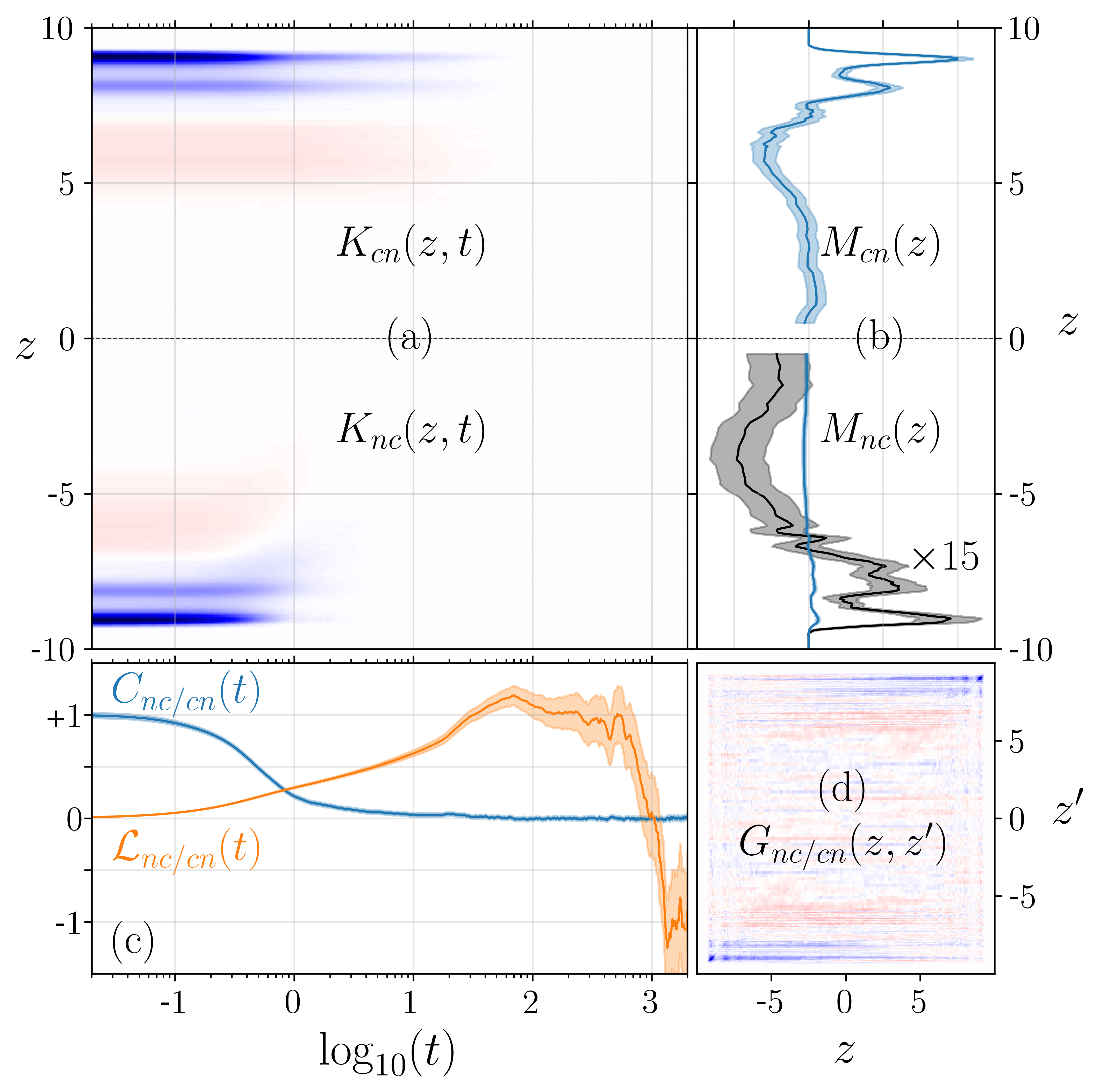}
    \end{minipage}
    
    \caption{\begin{small}
    \textbf{Off-diagonal fluid-charge kernels reveal interfacial electrokinetic coupling.}
    (\textbf{A}) Schematic representation of the two reciprocal electrokinetic responses. Electro-osmotic flow (EOF) corresponds to a particle-flux response to electrical driving, whereas streaming current (SC) corresponds to a charge-flux response to mechanical driving.
    (\textbf{B}) Response hierarchy for the mass--charge couplings. The layout follows Fig.~\ref{fig:diag_responses}C,D, but the reciprocal particle--charge components are shown together. Colors in the space--time and two-point kernels use a symmetric diverging scale, with blue and red indicating responses of opposite sign and white indicating zero.
    (a) Local--global kernels \(K_{nc}(z,t)\), corresponding to EOF, and \(K_{cn}(z,t)\), corresponding to SC.
    (b) Corresponding stationary profiles \(M_{nc}(z)\) and \(M_{cn}(z)\). The profile \(M_{nc}(z)\) is rescaled by a factor of 15 (black line) to compare spatial structure.
    (c) Global temporal projections \(C_{nc/cn}(t)\) and cumulative Green--Kubo integrals \(\mathcal{L}_{nc/cn}(t)\) that coincide following Onsager reciprocity.
    (d) Stationary two-point spatial response \(G_{nc/cn}(z,z')\), illustrating the nonlocal interfacial structure of electrokinetic coupling.
    \end{small}}
    \label{fig:eos_response}
\end{figure}

We now turn to the off-diagonal fluid--charge responses associated with streaming current (SC), \(R_{cn}(z,z',t)\), and electro-osmotic flow (EOF), \(R_{nc}(z,z',t)\). These two responses are related by Onsager reciprocity after proper projection, but their local--global kernels reveal distinct dynamical signatures.

Starting from the interface, the spatiotemporal kernels \(K_{cn}(z,t)\) and \(K_{nc}(z,t)\) (Fig.~\ref{fig:eos_response}B(a), top and bottom, respectively) show a region of strong positive correlations followed by a diffuse negative region that decays toward the pore mid-plane. At short times, the spatial organization of both kernels broadly follows the charge distribution in Fig.~\ref{fig:diag_responses}B. Their temporal evolution, however, differs markedly. The SC kernel relaxes in a comparatively homogeneous and slow way across the interfacial region, as shown by the nearly parallel contours. By contrast, the EOF kernel departs from this behavior around \(t\sim0.2\), where the correlation contours bend toward the pore center. This reflects two distinct local dynamical pathways (Fig.~\ref{fig:eos_response}A): SC is associated with the collective motion of the EDL driven by the fluid, whereas EOF involves local forcing of interfacial charge and subsequent nonlocal redistribution of momentum through viscous relaxation.

As a consequence, the SC profile \(M_{cn}(z)\) (Fig.~\ref{fig:eos_response}B(b), top), closely follows the charge-density distribution within the pore. In contrast, the EOF profile \(M_{nc}(z)\) (Fig.~\ref{fig:eos_response}B(b), bottom) has a peak amplitude about an order of magnitude smaller than its reciprocal counterpart. Its spatial features also differ. In particular, the negative diffuse contribution extends farther into the pore center, revealing a backflow-like contribution in the central region. This indicates that the local EOF and SC follow distinct local pathways, even though their fully integrated responses are constrained by Onsager reciprocity.

The temporal projections \(C_{nc/cn}(t)\) and \(\mathcal{L}_{nc/cn}(t)\) (Fig.~\ref{fig:eos_response}B(c)) further highlight the complexity of the off-diagonal responses. Although the long-time tail is only weakly visible in the raw correlation function \(C_{nc/cn}(t)\), the integrated response \(\mathcal{L}_{nc/cn}(t)\) exhibits clear structure over a much wider range of time scales than the diagonal responses \(\mathcal{L}_{nn}(t)\) and \(\mathcal{L}_{cc}(t)\) shown in Fig.~\ref{fig:diag_responses}. This shows that, even when the global hydrodynamic response contributes negligibly to the total fluid flux at \(t>70\) (Fig.~\ref{fig:diag_responses}C(c)), differences in hydrodynamic relaxation at different distances from the surface can still contribute to the net charge response up to \(t>1000\). These competing contributions are absent from purely steady-state descriptions and become visible only through the time-resolved kernel.

This behavior also highlights that electrokinetic response cannot be inferred from the net charge alone. Although the fluid is globally enriched in positive counterions, the fully integrated electrokinetic response is negative. The sign of the response therefore reflects not only the total charge content, but also the spatial organization of the EDL, the local mobility of the charged species, and the dynamical weighting imposed by hydrodynamic relaxation.

Finally, the stationary two-point response \(G_{nc/cn}(z,z')\) (Fig.~\ref{fig:eos_response}B(d)) confirms that electrokinetic coupling is interfacial but not purely local. Correlations extend along the charged layers and couple different distances from the walls, indicating that the response cannot be reduced to a local product of charge density and mobility. Reciprocity is recovered at this level in its proper nonlocal form, \(G_{ab}(z,z')=G_{ba}(z',z)\). Apparent differences between the projected profiles \(M_{ab}(z)\) and \(M_{ba}(z)\) therefore reflect one-sided spatial projection of a nonlocal response, not a violation of Onsager symmetry, as discussed in the Supplementary Text.

\subsection*{The full response matrix reveals transport pathways across all channels}
\label{subsec:full}
\begin{figure}
    \centering
    \begin{minipage}[t]{0.60\textwidth}
        \centering
        {\Large\bfseries A}\\[0.1em]
        \includegraphics[width=\textwidth]{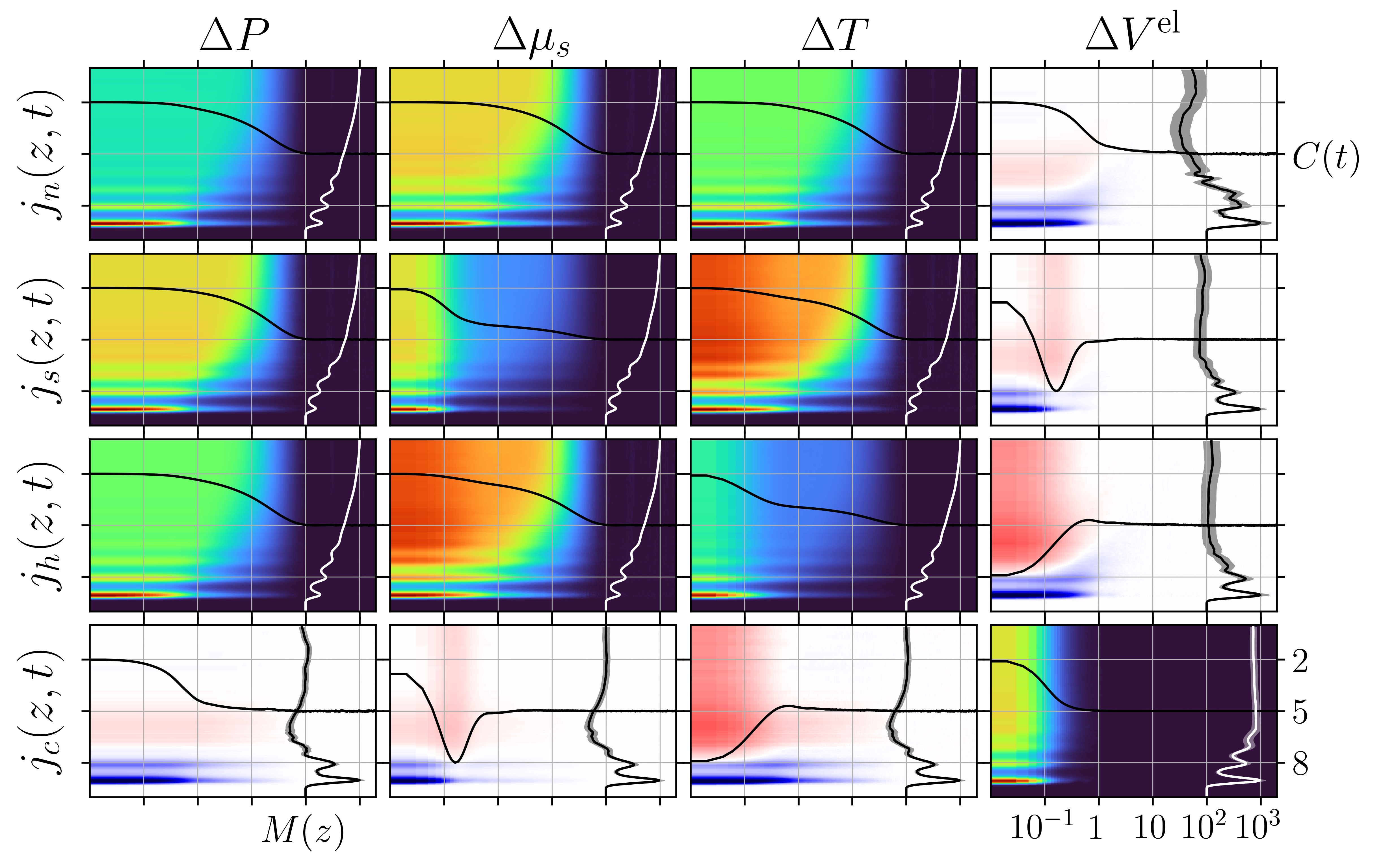}
    \end{minipage}
    \hfill
    \begin{minipage}[t]{0.38\textwidth}
        \centering
        {\Large\bfseries B}\\[0.1em]
        \includegraphics[width=1.\textwidth]{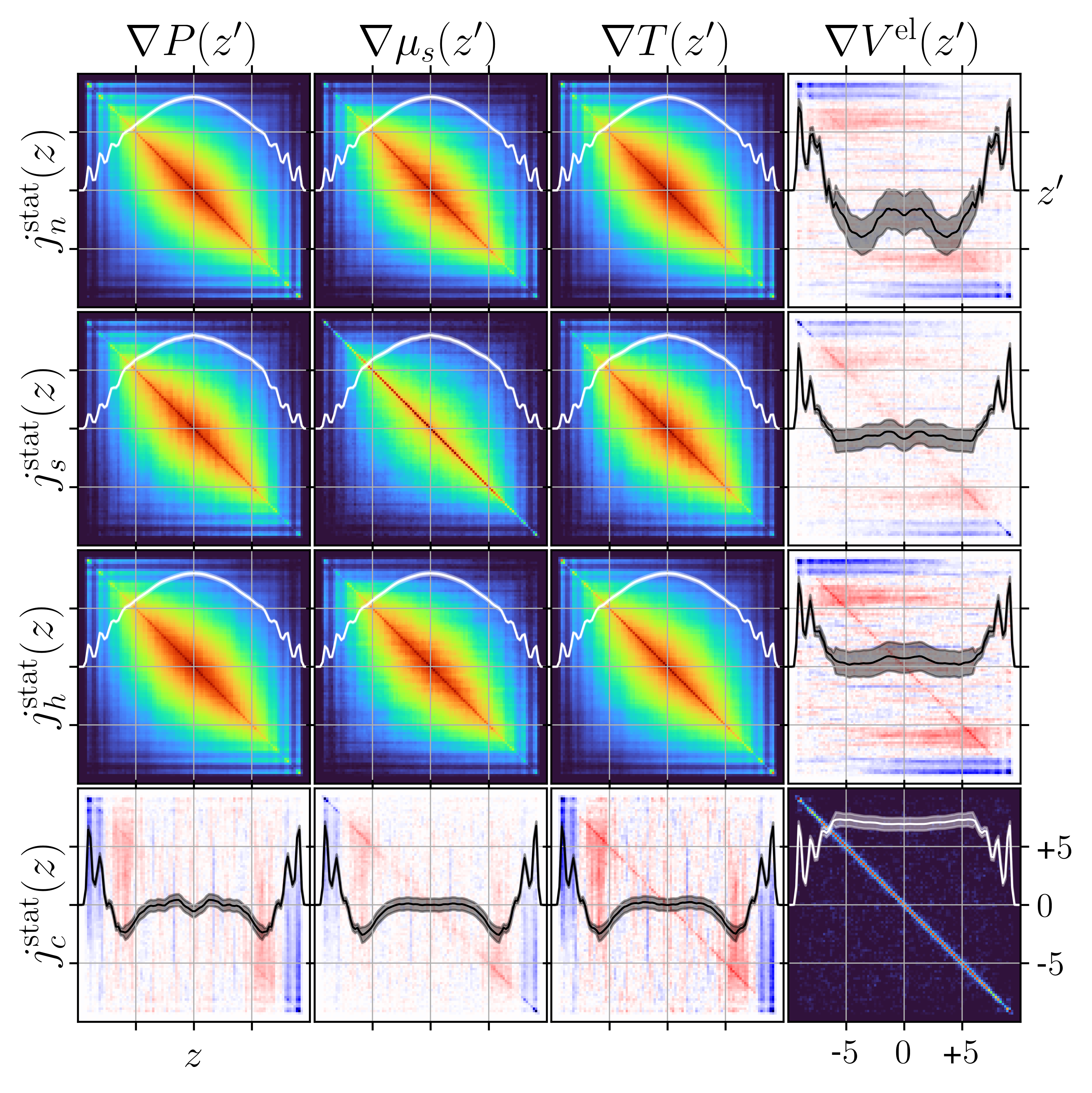}
    \end{minipage}
    \caption{\begin{small}
    \textbf{Spatio-temporal view of fluid--solute--heat--charge coupled transport in confined fluids.}
    (\textbf{A}) Matrix representation of the sixteen reduced response kernels \(K_{ab}(z,t)=\int R_{ab}(z,z',t)\,\mathrm{d}z'\), where rows denote the measured flux channel \(a\in\{n,s,h,c\}\) and columns denote the conjugate driving channel \(b\in\{n,s,h,c\}\). Each panel shows the local--global response as a function of position \(z\) and logarithmic time \(t\), together with the corresponding normalized stationary profile \(M_{ab}(z)\) and global time correlation \(C_{ab}(t)\). (\textbf{B}) Matrix representation of the stationary local--local responses \(G_{ab}(z,z')=\int_0^\infty R_{ab}(z,z',t)\,\mathrm{d}t\). Each panel shows the spatial correlation between observation position \(z\) and source position \(z'\), together with the normalized projection \(M_{ab}(z)=\int G_{ab}(z,z')\,\mathrm{d}z'\).  An excess-flux representation of the solute and heat channels is discussed in the Supplementary Text.
    \end{small}}
  \label{fig:full_matrix}
\end{figure}

The results presented above illustrate how the response-kernel framework turns equilibrium fluctuations into a transparent diagnostic of coupled electrokinetic transport. This pair of reciprocal responses, however, represents only one sector of a larger object.  We therefore extend the analysis to the full \(4\times4\) matrix of reduced space--time response kernels coupling particle (\(n\)), solute (\(s\)), heat (\(h\)), and charge (\(c\)) transport to their conjugate driving forces. Figure~\ref{fig:full_matrix}A displays the sixteen local--global kernels \(K_{ab}(z,t)\), together with their stationary projections \(M_{ab}(z)\) and global temporal projections \(C_{ab}(t)\). Figure~\ref{fig:full_matrix}B shows the complementary stationary local--local responses \(G_{ab}(z,z')\).

The full response matrix reveals a clear block structure. Particle, solute, and heat transport form a hydrodynamically dominated sector: the corresponding \(3\times3\) submatrix displays long-lived relaxation patterns in \(K_{ab}(z,t)\) together with extended spatial correlations in \(G_{ab}(z,z')\). This common structure shows that solute and heat transport remain strongly coupled to fluid motion through advective and collective relaxation processes (see Supplementary Text). At the same time, the solute and heat channels are not purely hydrodynamic. Already at the diagonal level, the \(ss\) and \(hh\) responses combine a short-time relaxation associated with local solute, ionic, or thermal fluctuations and a slower pore-scale relaxation governed by hydrodynamics.

Several physical insights follow from the representation of coupled transport developed in this work. Momentum transport is governed by long-lived, nonlocal hydrodynamic modes extending across the pore, whereas charge transport relaxes rapidly and locally through ionic friction and electrostatic screening~\cite{bazant2004diffuse}. Solute and heat transport occupy an intermediate position: their stationary profiles are strongly affected by hydrodynamic relaxation, while their time-dependent kernels also reveal faster microscopic contributions. Off-diagonal electrokinetic, diffusio-electric, and thermo-electric couplings combine these sectors, inheriting interfacial localization from molecular structure while coupling to slower collective relaxation.

The full matrix analysis shows what becomes accessible when molecular transport is treated as a field-level response instead of being immediately compressed into coefficients, excess quantities, or boundary conditions. In this sense, quantities such as slip lengths, zeta potentials, surface conductivities, Kapitza resistances, interfacial dielectric responses, and liquid--solid friction coefficients can be viewed as different coarse-grained summaries of interfacial molecular dynamics. Their apparent locality may hide nonlocal electrostatic, hydrodynamic, electronic, and correlation effects~\cite{Underwood2022dielectric,duque_2020_nonlocal}, as illustrated by recent work on water--carbon interfaces, where liquid charge fluctuations can couple to electronic excitations in the confining wall and produce quantum or electronic contributions to interfacial friction~\cite{kavokine2022fluctuations,Bui2023quantum}. This perspective is closely related to recent efforts to clarify how Green--Kubo expressions for liquid--solid friction are connected to effective hydrodynamic boundary conditions and slip lengths~\cite{Bui2024friction}.

This perspective also clarifies the status of reciprocity in inhomogeneous systems. Onsager symmetry is naturally expressed at the level of the nonlocal response kernel, where both source and observation positions are retained. Projected quantities such as \(K_{ab}(z,t)\), \(M_{ab}(z)\), or effective interfacial coefficients contain only part of this spatial information and need not display a pointwise symmetry between reciprocal channels. This is not a violation of microscopic reversibility, but a consequence of projection and of the chosen flux--force representation.

More broadly, the response-kernel formulation suggests a route beyond models that treat molecular effects only as corrections to bulk hydrodynamics. Because the kernel retains the space, time, and cross-coupling structure of the microscopic response, it provides natural targets for physics-informed machine learning and improved estimators~\cite{sammuller2023neural}. It is complementary to mesoscale approaches such as mode-coupling theory, dynamic density functional theory, and fluctuating hydrodynamics, which also encode collective modes, memory, density correlations, or stochastic hydrodynamic fields~\cite{Grmela1997dynamics,das_MCT_2004,donev2019fhd,schmidt2022PFT,obliger2023simple,schlaich2025theory}. It therefore offers a microscopic basis for constitutive laws that retain nonlocality, relaxation, and cross-coupling before reducing them to a continuum closure.

Several limitations and challenges should be kept in view, as discussed in the Supplementary Text. Although the framework presented here is demonstrated on a minimal model fluid, without the chemical specificity, polarization, or reactivity of real interfaces, the most fundamental challenge is to connect microscopic response functions to the Onsager matrices measured experimentally. In experiments, reservoir-imposed differences in pressure, temperature, composition, and electric potential generate coupled perturbations rather than independent local forces. Projecting mechanically defined microscopic currents onto experimental control variables, local thermodynamic potentials, and effective boundary conditions therefore remains a central coarse-graining problem.

A major opportunity lies in connecting molecular transport theory with the design of functional fluidic systems. Confined fluids may require a conceptual shift from a few effective coefficients toward response functions that retain molecular structure, interfacial dynamics, and coupled transport pathways. This ambition is not purely speculative: biological membranes, ion channels, and nanopores already exploit spatially organized coupled transport, combining molecular selectivity, electrostatic coupling, conformational dynamics, chemical activity, and nonequilibrium driving with remarkable efficiency. Recent developments in field-controlled and iontronic nanofluidics suggest that artificial systems are beginning to exploit related mechanisms~\cite{Bui2026dielectrocapillarity,kamsma2024brain}. Although the present work remains a minimal model fluid, it points toward a response-based description of soft and charged interfaces in which relaxation and coupling become microscopic design variables for applications in nanofluidics, iontronics, energy conversion, chemical separation, sensing, and biomedical transport.


\clearpage 

\bibliography{bibliography} 
\bibliographystyle{sciencemag}

%
%
%
%
%
%


\newpage

\section*{Acknowledgments}
We thank Princeton Research Computing for technical support and access to shared computational resources.

\paragraph*{Funding:}
This work was supported by the US Department of Energy, Office of Science, Office of Basic Energy Sciences, Chemical Sciences, Geosciences, and Biosciences Division, Geosciences Program under Award DE-SC0018419. Computational resources were provided by Princeton University's shared Research Computing facilities.

\paragraph*{Disclaimer:}
This manuscript was prepared as an account of work sponsored by an agency of the United States Government. Neither the United States Government nor any agency thereof, nor any of their employees, makes any warranty, express or implied, or assumes any legal liability or responsibility for the accuracy, completeness, or usefulness of any information, apparatus, product, or process disclosed, or represents that its use would not infringe privately owned rights. Reference herein to any specific commercial product, process, or service by trade name, trademark, manufacturer, or otherwise does not necessarily constitute or imply its endorsement, recommendation, or favoring by the United States Government or any agency thereof. The views and opinions of authors expressed herein do not necessarily state or reflect those of the United States Government or any agency thereof.

\paragraph*{Author contributions:}
M.-T.H.N. designed the study, performed the simulations, analyzed the data, and prepared the figures. I.C.B. contributed to conceptual development, supervision, and interpretation of the results. Both authors discussed the results and wrote the manuscript.

\paragraph*{Competing interests:}
There are no competing interests to declare.

\paragraph*{Data, code and materials availability:}
The simulation data, simulation input files, and analysis scripts underlying this work will be deposited in a public repository before publication and made available without embargo. No physical materials were generated in this work.

\newpage

\subsection*{Supplementary materials}
Materials and Methods\\
Supplementary Text\\
Figures S1 to S2\\
Tables S1 to S2\\
References \textit{(47-\arabic{enumiv})}\\ 

\newpage


\renewcommand{\thefigure}{S\arabic{figure}}
\renewcommand{\thetable}{S\arabic{table}}
\renewcommand{\theequation}{S\arabic{equation}}
\renewcommand{\thepage}{S\arabic{page}}
\setcounter{figure}{0}
\setcounter{table}{0}
\setcounter{equation}{0}
\setcounter{page}{1} 


\begin{center}
\section*{Supplementary Materials for\\ \scititle}

Minh-Thê Hoang Ngoc$^{\ast}$,
Ian C. Bourg\\ 
\small$^\ast$Corresponding author. Email: mh7003@princeton.edu\\
\end{center}

\subsubsection*{This PDF file includes:}
Materials and Methods\\
Supplementary Text\\
Figures S1 to S2\\
Tables S1 to S2\\


\subsection*{Materials and Methods}\label{materials_methods}

\subsubsection*{Model charged nanopore}\label{materials_methods:system}

We consider a minimal molecular model of a charged electrolyte confined between two solid walls. All quantities are reported in reduced LJ units, with particle diameter \(\sigma\), energy scale \(\epsilon\), particle mass \(m\), and time unit \(\tau=\sigma\sqrt{m/\epsilon}\) (\(\sigma=\epsilon=m=e=k_B=1\)). The simulation cell has lateral dimensions \(L_x=L_y=50\sigma\), with periodic boundary conditions in the lateral directions. The two solid walls are separated by a distance \(H=20\sigma\), forming a planar slit nanopore.

The confined fluid contains \(N_f=30976\) mobile particles, composed of \(22976\) neutral solvent particles, \(4250\) cations, and \(3750\) anions. Cations and anions carry charges \(q=\pm 1\), respectively. The walls are represented by two planes of solid atoms arranged on a square lattice of spacing \(\sigma\), with \(2500\) atoms per wall. Each solid atom is tethered to a fixed ghost atom at its lattice position by a harmonic spring with equilibrium distance zero and is also harmonically bonded to its four nearest neighbors in the wall plane. All harmonic springs in the system have spring constant \(k=100,\epsilon/\sigma^2\), allowing the walls to fluctuate thermally around their reference crystalline structure.

The wall atoms carry a partial charge \(q_s=-0.1\), producing an overall charged interface. Pair interactions consist of LJ, Weeks--Chandler--Andersen (WCA), Coulomb, and harmonic contributions, depending on the chemical species. The full interaction matrix is summarized in Table~\ref{sup_tab:interaction_matrix}. Lennard--Jones interactions are truncated at \(r_c=2.5\sigma\), while WCA interactions are truncated at \(r_c=2^{1/6}\sigma\).

\begin{table}[!htbp]
\centering
\begin{tabular}{c|cccc}
\textbf{} & \textbf{N} & \textbf{+} & \textbf{-} & \textbf{S} \\
\hline
\textbf{N} & LJ & LJ & WCA & LJ \\
\textbf{+} & -- & LJ + C (rep.) & WCA + C (att.) & LJ + C (att.) \\
\textbf{-} & -- & -- & LJ + C (rep.) & WCA + C (rep.) \\
\textbf{S} & -- & -- & -- & H + C (rep.)\\
\end{tabular}
\caption{\textbf{Pair interactions in the model charged nanopore.}
N denotes neutral solvent particles, \(+\) cations, \(-\) anions, and S solid wall atoms. LJ denotes a Lennard--Jones interaction, WCA a purely repulsive Weeks--Chandler--Andersen interaction, C a Coulomb interaction, and H harmonic bonding or tethering. Attractive and repulsive Coulomb interactions are indicated according to the signs of the interacting charges.}
\label{sup_tab:interaction_matrix}
\end{table}

\subsubsection*{Local densities and currents}\label{materials_methods:observables}

We resolve transport through local densities and currents associated with four transported quantities: total particle number \(n\), solute number \(s\), heat \(h\), and charge \(c\), following the Irving--Kirkwood construction of microscopic hydrodynamic fields~\cite{yang2012generalized}. For a microscopic configuration \(\Gamma(t)\), the corresponding density fields are defined as
\begin{align}
    \rho_n(\mathbf{r},t) 
    &= \sum_i \delta(\mathbf{r}-\mathbf{r}_i(t)), \\
    \rho_s(\mathbf{r},t) 
    &= \sum_i \chi_i^s \delta(\mathbf{r}-\mathbf{r}_i(t)), \\
    \rho_h(\mathbf{r},t) 
    &= \sum_i e_i(t)\delta(\mathbf{r}-\mathbf{r}_i(t)), \\
    \rho_c(\mathbf{r},t) 
    &= \sum_i q_i\delta(\mathbf{r}-\mathbf{r}_i(t)),
\end{align}
where \(\chi_i^s=1\) for solute particles and \(0\) otherwise, \(q_i\) is the particle charge, and \(e_i\) is the single-particle energy. The associated microscopic currents are
\begin{align}
    \mathbf{j}_n(\mathbf{r},t) 
    &= \sum_i \mathbf{v}_i(t)\delta(\mathbf{r}-\mathbf{r}_i(t)), \\
    \mathbf{j}_s(\mathbf{r},t) 
    &= \sum_i \chi_i^s\mathbf{v}_i(t)\delta(\mathbf{r}-\mathbf{r}_i(t)), \\
    \mathbf{j}_h(\mathbf{r},t) 
    &= \sum_i e_i(t)\mathbf{v}_i(t)\delta(\mathbf{r}-\mathbf{r}_i(t))
    -
    \frac{1}{2}\sum_{i,j}
    \left[
    \mathbf{F}_{ij}(t)\cdot\mathbf{v}_i(t)
    \right]
    \mathbf{r}_{ij}(t)\,
    \delta(\mathbf{r}-\mathbf{r}_i(t)), \\
    \mathbf{j}_c(\mathbf{r},t) 
    &= \sum_i q_i\mathbf{v}_i(t)\delta(\mathbf{r}-\mathbf{r}_i(t)).
\end{align}
In practice, the heat/energy current was computed using a standard Irving--Kirkwood--Hardy-type microscopic energy-current expression, including kinetic, potential, and virial contributions~\cite{ferreira2026consistency}. Because the spatial localization of interaction-energy transport is not unique at the molecular scale, the heat-related kernels should be interpreted as local resolutions defined within this specified microscopic-current convention. For analysis, the microscopic currents are laterally averaged over bins along the confinement direction. For a bin \(p\) centered at \(z_p\) with width \(\Delta z_p\), we define
\begin{equation}
    \mathbf{j}_a^{p}(t)
    =
    \frac{1}{S}
    \int_{z_p-\Delta z_p/2}^{z_p+\Delta z_p/2}
    \mathbf{j}_a(\mathbf{r},t)\,\mathrm{d}^3\mathbf{r},
\end{equation}
where \(S=L_xL_y\) is the lateral surface area. The pore was discretized into \(260\) nonuniform bins along \(z\), with \(\Delta z=0.04\sigma\) near the interfaces and \(\Delta z=0.2\sigma\) in the pore center. For the particle, solute, and charge currents, this lateral binning can be written explicitly as
\begin{equation}
    \mathbf{j}_a^{p}(t)
    =
    \frac{1}{S}
    \sum_i A_i^a(t)\mathbf{v}_i(t)
    \Theta\left(\frac{\Delta z_p}{2}-|z_i(t)-z_p|\right),
    \qquad
    a\in\{n,s,c\},
\end{equation}
with \(A_i^n=1\), \(A_i^s=\chi_i^s\), and \(A_i^c=q_i\). The heat current contains both convective and interaction/virial contributions and was binned using the corresponding microscopic heat-current convention defined above.

Finally, for the slab geometry of the simulated system, we project the three-dimensional response onto longitudinal fluxes resolved across the confinement direction:
\begin{equation}
    j_a^x(z,t)
    =
    \frac{1}{S}
    \int_S
    \vec{j}_a(\vec r,t)\cdot\hat{\vec e}_x
    \, \mathrm{d}^2S ,
    \label{eq:local_flux_slab}
\end{equation}
where \(S\) is the area parallel to the walls. The corresponding reduced kernel is
\begin{equation}
    R_{ab}(z,z',t)
    =
    \frac{1}{k_B T}
    \left\langle
    j_a^x(z,t) j_b^x(z',0)
    \right\rangle_\mathrm{eq},
    \label{eq:R_zz_t}
\end{equation}
with \(a,b\in\{n,s,h,c\}\), denoting fluid-particle, solute, heat, and charge transport. This reduced kernel remains a matrix of space--time response functions, but is adapted to the symmetry of the confined system.

\subsubsection*{Molecular dynamics simulations}\label{materials_methods:simu}

Molecular dynamics simulations were performed using the program LAMMPS~\cite{LAMMPS}. The system was evolved in the canonical ensemble at temperature \(T=1.4\), maintained with a Nosé--Hoover thermostat with damping time \(0.1\). The equations of motion were integrated with a time step \(\Delta t=10^{-3}\). Long-range electrostatic interactions were computed using the particle-particle, particle-mesh (PPPM) algorithm with a relative precision of \(10^{-5}\), together with the Yeh--Berkowitz slab correction using a vacuum spacing factor of 3.0.

Initial configurations were prepared by placing the mobile particles in the confined region and assigning velocities from a Maxwell--Boltzmann distribution at the target temperature. Each trajectory was equilibrated for \(10^6\) time steps, corresponding to a reduced time \(t=1000\). Production trajectories were then run for \(4\times10^6\) time steps, corresponding to \(t=4000\). Local currents and densities were sampled every 20 time steps.

To improve ensemble statistics, the full simulation protocol was repeated over 256 independent replicas, initialized with different random seeds for the ion positions and particle velocities. The two lateral directions \(x\) and \(y\) were treated as statistically equivalent independent samples when computing transport correlations.

\subsubsection*{Extraction of response kernels}\label{materials_methods:analysis}

The central quantities of this work are equilibrium current correlations resolved in space, time, and transported variable. The most complete response object is the nonlocal kernel \(R_{ab}(z,z',t)\), which relates the current of transported quantity \(a\) at position \(z\) and time \(t\) to the conjugate driving force associated with quantity \(b\) at source position \(z'\) and initial time. Directly storing and analyzing the full time-dependent tensor \(R_{ab}(z,z',t)\) is computationally demanding. We therefore use two complementary projections. The local--global kernels \(K_{ab}(z,t)\) retain the time dependence of the response and are computed directly from Green--Kubo current correlations. The stationary local--local kernels \(G_{ab}(z,z')\) retain the two-position structure and are extracted using associated Einstein--Helfand relations~\cite{Erpenbeck1995einstein}.

We first compute local--global response kernels of the form
\begin{equation}
    K_{ab}(z_p,t)
    =
    \left\langle j_a^p(t) J_b(0) \right\rangle,
    \label{eq:Kab_definition_methods}
\end{equation}
where \(j_a^p(t)\) is the laterally averaged current of quantity \(a\) in bin \(p\), and
\begin{equation}
    J_b(t)
    =
    \sum_p j_b^p(t)\Delta z_p
    \label{eq:Jb_definition_methods}
\end{equation}
is the corresponding global current across the pore. Correlation functions were computed from the equilibrium current time series using fast Fourier transforms. Ensemble averages were performed over the 256 independent replicas and over the two equivalent lateral directions. All uncertainty bands and reported errors correspond to approximate 95\% confidence intervals, estimated as \(\pm 2\,\mathrm{SE}\), where \(\mathrm{SE}\) is the standard error over the ensemble samples.

Stationary local transport profiles were obtained by time integration,
\begin{equation}
    M_{ab}(z_p)
    =
    \int_0^\infty K_{ab}(z_p,t)\,\mathrm{d}t,
\end{equation}
with the upper integration time chosen after convergence of the corresponding running integral. The global time-dependent response was obtained by spatial integration,
\begin{equation}
    C_{ab}(t)
    =
    \int_{-H/2}^{H/2} K_{ab}(z,t)\,\mathrm{d}z.
\end{equation}

We also compute stationary local--local response functions,
\begin{equation}
    G_{ab}(z_p,z_q)
    =
    \int_0^\infty
    \left\langle j_a^p(t)j_b^q(0)\right\rangle
    \mathrm{d}t.
\end{equation}
Instead of explicitly storing the full time-dependent tensor, these stationary two-point responses were evaluated using an Einstein--Helfand representation. We define the integrated current
\begin{equation}
    Q_a^p(t)
    =
    \int_0^t j_a^p(s)\,\mathrm{d}s.
    \label{eq:Helfand_moment_methods}
\end{equation}
The stationary response is then obtained from the long-time growth rate of the Helfand covariance,
\begin{equation}
    G_{ab}(z_p,z_q)
    =
    \frac{1}{2}
    \lim_{t\to\infty}
    \frac{\mathrm{d}}{\mathrm{d}t}
    \left\langle
    \Delta Q_a^p(t)\Delta Q_b^q(t)
    \right\rangle,
    \label{eq:Gab_Helfand_methods}
\end{equation}
where \(\Delta Q_a^p(t)=Q_a^p(t)-Q_a^p(0)\). In practice, the slope was extracted by fitting the linear regime of the growth; for all responses analyzed here, this regime was reached for \(t>1000\).

\begin{small}
\begin{table}
\centering
\begin{tabular}{c|cccc}
 & $\cell{\phi_n = \frac{V}{N}P \\ \text{pressure}}$
 & $\cell{\phi_s=\mu_s \\ \text{chemical pot.}}$
 & $\cell{\phi_h = T/T_0 \\ \text{temperature}}$
 & $\cell{\phi_c= V^\mathrm{el} \\ \text{electric pot.}}$ \\
\hline
$\cell{n\ [1] \\ \text{particle}}$
& $(1.218\!\pm\!0.014)\!\times\!\!10^{1}$
& $(3.39\!\pm\!0.04)\!\times\!\!10^{0}$
& $(1.13\!\pm\!0.01)\!\times\!\!10^{2}$
& $(-1.9\!\pm\!0.5)\!\times\!\!10^{-2}$ \\

$\cell{s\ [1] \\ \text{solute}}$
& $(3.39\!\pm\!0.05)\!\times\!\!10^{0}$
& $(9.7\!\pm\!0.1)\!\times\!\!10^{-1}$
& $(3.14\!\pm\!0.04)\!\times\!\!10^{1}$
& $(-9.7\!\pm\!1.6)\!\times\!\!10^{-3}$ \\

$\cell{h\ [\mathrm{energy}] \\ \text{heat}}$
& $(1.13\!\pm\!0.01)\!\times\!\!10^{2}$
& $(3.14\!\pm\!0.04)\!\times\!\!10^{1}$
& $(1.06\!\pm\!0.01)\!\times\!\!10^{3}$
& $(-2.4\!\pm\!0.4)\!\times\!\!10^{-1}$ \\

$\cell{c\ [\mathrm{charge}] \\ \text{charge}}$
& $(-1.8\!\pm\!2.8)\!\times\!\!10^{-2}$
& $(-9.5\!\pm\!7)\!\times\!\!10^{-3}$
& $(-2.3\!\pm\!2.6)\!\times\!\!10^{-1}$
& $(2.03\!\pm\!0.12)\!\times\!\!10^{-2}$
\end{tabular}
\caption{
Numerical values of the volume-normalized Onsager matrix, \(L_{ab}/V\), obtained from equilibrium molecular dynamics of the confined charged Lennard--Jones fluid. The coefficients are computed using \(L_{ab}=(k_\mathrm{B}T)^{-1}\int_0^\infty \left\langle J_a(t)J_b(0)\right\rangle\,\diff t\), and enter the linear response relation \(J_a=-\sum_b L_{ab}\nabla\delta\phi_b\), where \([J_a]=[a]\,\mathrm{length}/\mathrm{time}\) and \([\phi_a]=\mathrm{energy}/[a]\). The reported values are normalized by the simulation control volume \(V=50\times50\times20\,\sigma^3\), so that \([L_{ab}/V]=[a][b]/(\mathrm{energy}\,\mathrm{time}\,\mathrm{length})\). In Lennard--Jones units, the numerical values are therefore given in \([a][b]/(k_\mathrm{B}T\,\sigma\,\tau)\).
}
\label{sup_tab:L_ab_numerical}
\end{table}
\end{small}

\subsubsection*{Analytical reference models}\label{materials_methods:analytical}

\begin{figure}
    \centering
    \includegraphics[width=.5\textwidth]{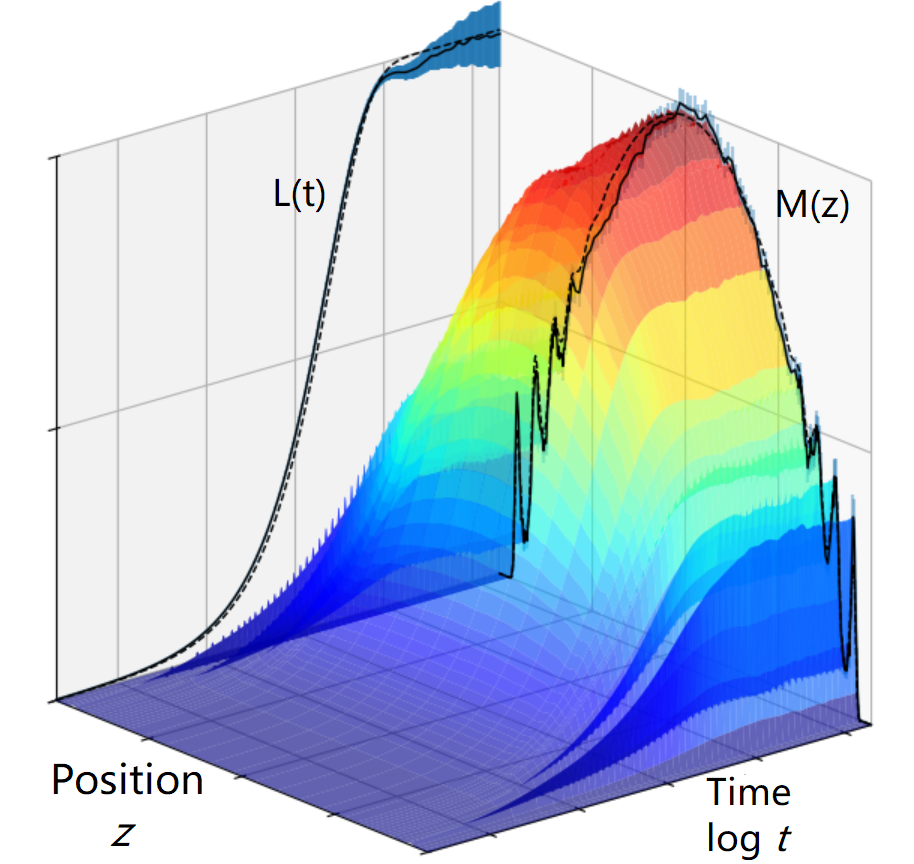}
    \caption{\begin{small}
    \textbf{Fitting hydrodynamic parameters from time-dependent response.} Fluid response \(\mathcal{M}_{nn}(z,t)\), shown together with the corresponding stationary profile \(M_{nn}(z)\) and global running integral \(\mathcal{L}_{nn}(t)\). Dashed lines indicate the best fit to the unsteady Stokes model, from which the effective confinement and viscosity are extracted.
\end{small}}
    \label{sup_fig:fit_stokes}
\end{figure}

To interpret the diagonal particle and charge responses, we compare the response kernels extracted from MD simulations with two minimal reference models. These models are not intended to provide an exhaustive continuum description of the confined fluid. Rather, they illustrate how the response-kernel framework can be connected, after appropriate projections, to familiar physical pictures and how additional information becomes available when the full space--time structure is retained. The fluid response is described by the Green function of the unsteady Stokes equation in a planar no-slip slit, which captures the nonlocal and long-lived character of momentum transport across the pore. The charge response is interpreted using a simpler local Nernst--Einstein picture, appropriate when ionic conduction is dominated by local ionic mobility and density. These two reference descriptions are compared with the measured responses in Fig.~\ref{fig:diag_responses}C,D and Fig.~\ref{sup_fig:fit_stokes}.

\paragraph*{Hydrodynamic reference model}

The hydrodynamic reference model is based on the unsteady Stokes equation
\begin{equation}
    \left(\partial_t-\nu\partial_z^2\right)u(z,t)
    =
    -\frac{1}{\rho_n^b}\frac{\partial p}{\partial x}(z,t),
    \qquad
    u(\pm H/2,t)=0,
    \qquad
    u(z,0)=0,
    \label{eq:unsteady_stokes}
\end{equation}
where \(\nu\) is the kinematic viscosity and
\(\rho_n^b=\langle\rho_n(0)\rangle\) is the reference bulk density at the center of the pore. The particle current is related to the velocity field by
\begin{equation}
    j_n(z,t)=\langle\rho_n(z)\rangle u(z,t).
\end{equation}
The nonlocal response kernel \(R_{nn}(z,z',t)\) is identified with the Green function relating a localized mechanical forcing at \(z'\) to the induced particle flux at \(z\). The local--global response \(K_{nn}(z,t)\), the running and stationary local--local responses \(\mathcal{G}_{nn}(z,z',t)\) and \(G_{nn}(z,z')\), and the corresponding spatially integrated quantities are then obtained by projection of this Green function.

For the unsteady Stokes model, the local--local particle response kernel is
\begin{equation}
    R_{nn}(z,z',t)
    =
    \frac{2\langle\rho_n(z)\rangle}{H}
    \sum_{m=1}^{\infty}
    \varphi_m(z)\varphi_m(z')
    e^{-\nu k_m^2t},
    \label{eq:Rnn_stokes}
\end{equation}
where the no-slip eigenmodes are
\begin{equation}
    \varphi_m(z)=\sin\!\left[k_m(z+H/2)\right],
    \qquad
    k_m=\frac{m\pi}{H}.
\end{equation}

The corresponding running local--local response,
\(\mathcal{G}_{nn}(z,z',t)=\int_0^t R_{nn}(z,z',s)\,\mathrm{d}s\), is
\begin{equation}
    \mathcal{G}_{nn}(z,z',t)
    =
    \frac{2\langle\rho_n(z)\rangle H}{\nu\pi^2}
    \sum_{m=1}^{\infty}
    \varphi_m(z)\varphi_m(z')
    \frac{1-e^{-\nu k_m^2t}}{m^2}.
    \label{eq:Gnn_time_stokes}
\end{equation}
Taking the long-time limit gives the stationary local--local response
\begin{equation}
    G_{nn}(z,z')
    =
    \frac{\langle\rho_n(z)\rangle}{\nu H}
    \left(H/2+\min(z,z')\right)
    \left(H/2-\max(z,z')\right).
    \label{eq:Gnn_stokes}
\end{equation}

The local--global response, obtained by integrating over the source coordinate \(z'\), is
\begin{equation}
    K_{nn}(z,t)
    =
    \frac{4\langle\rho_n(z)\rangle}{\pi}
    \sum_{\substack{m=1\\ m~\mathrm{odd}}}^{\infty}
    \frac{\varphi_m(z)}{m}
    e^{-\nu k_m^2t},
    \label{eq:Knn_stokes}
\end{equation}
and its running time integral is
\begin{equation}
    \mathcal{M}_{nn}(z,t)
    =
    \frac{4\langle\rho_n(z)\rangle H^2}{\nu\pi^3}
    \sum_{\substack{m=1\\ m~\mathrm{odd}}}^{\infty}
    \frac{\varphi_m(z)}{m^3}
    \left[
    1-e^{-\nu k_m^2t}
    \right],
    \label{eq:Mnn_time_stokes}
\end{equation}
which converges at long times to
\begin{equation}
    M_{nn}(z)
    =
    \frac{H^2\langle\rho_n(z)\rangle}{8\nu}
    \left[
    1-\left(\frac{2z}{H}\right)^2
    \right].
    \label{eq:Mnn_stokes}
\end{equation}
This expression recovers the classical Hagen--Poiseuille parabolic profile, weighted by the stationary density profile. In the homogeneous limit, \(\langle\rho_n(z)\rangle=\rho_n^b\), one recovers the classical Poiseuille permeability, or equivalently the Darcy-law conductance of a planar slit,
\begin{equation}
    L_{nn}=\frac{\rho_n^b H^3}{12\nu}.
\end{equation}

In a conventional analysis, one would estimate the effective confinement and viscosity by fitting only the stationary mobility profile \(M_{nn}(z)\). Here, the time-dependent response \(\mathcal{M}_{nn}(z,t)\) provides a much stronger constraint, since it contains the full transient relaxation toward the stationary hydrodynamic response. Fitting the stationary profile \(M_{nn}(z)\) yields
\[
H_{\mathrm{eff}} = 19.5\pm0.1,
\qquad
\nu = 1.4\pm0.1,
\]
whereas fitting the full spatiotemporal response \(\mathcal{M}_{nn}(z,t)\) yields
\[
H_{\mathrm{eff}} = 19.49\pm0.02,
\qquad
\nu = 1.465\pm0.005.
\]
These results, shown in Fig.~\ref{sup_fig:fit_stokes}, illustrate that using the full space--time response substantially improves parameter estimation. In practice, the fit is performed over the entire spatiotemporal dataset rather than over a single stationary profile, which greatly reduces the uncertainty on the inferred hydrodynamic parameters. The resulting model captures the main features of the measured response, including the parabolic stationary profile and the slow relaxation toward the hydrodynamic steady state, as shown in Fig.~\ref{fig:diag_responses}C.

\paragraph*{Local Nernst--Einstein reference for charge transport}

The diagonal charge response is modeled using a local Nernst--Einstein reference picture. This choice reflects the qualitative behavior observed in the molecular response: charge-current correlations relax rapidly and remain spatially localized, in contrast with the long-lived nonlocal relaxation of momentum.

In this reference description, the charge current is approximately Ohmic,
\begin{equation}
    j_c(z,t) \simeq \sigma_{\rm NE}(z) E(z,t),
\end{equation}
with a local Nernst--Einstein conductivity
\begin{equation}
    \sigma_{\rm NE}(z)
    =
    \frac{e^2}{k_{\rm B}T}
    \left[D_+ \rho_+(z)+D_- \rho_-(z)\right].
\end{equation}
For a symmetric electrolyte with comparable ionic diffusivities,
\(D_+ \simeq D_- \equiv D_{\rm ion}\), this reduces to
\begin{equation}
    \sigma_{\rm NE}(z)
    \simeq
    \frac{e^2 D_{\rm ion}}{k_{\rm B}T}\rho_s(z),
    \label{eq:sigma_NE_local}
\end{equation}
where \(\rho_s(z)=\rho_+(z)+\rho_-(z)\) is the local salt density. Conversely, the measured stationary charge response can be expressed as an effective local ionic diffusivity,
\begin{equation}
    D_{\rm ion}^{\rm eff}(z)
    =
    \frac{k_{\rm B}T}{e^2\rho_s(z)}\sigma(z),
    \label{eq:Dion_eff}
\end{equation}
where \(\sigma(z)\) denotes the local conductivity inferred from the diagonal charge response.

This comparison is intended as a local interpretive reference rather than as a full continuum theory of charge relaxation. The near agreement between the measured charge response and the Nernst--Einstein form indicates that ionic conduction in the model is close to ideal, with deviations mainly reflecting interfacial density structure and local mobility variations, as shown in Fig.~\ref{fig:diag_responses}D. This interpretation is also supported by the fast exponential relaxation of the charge response and by its comparatively uniform behavior across the pore. The absence of an extended hydrodynamic tail is consistent with the dominance of ionic friction and electrostatic screening over pore-scale momentum diffusion in the charge channel.

\subsubsection*{Linear response theory for Hamiltonian dynamics at the field level}\label{sup_text:LRT}
For completeness, we briefly summarize the derivation of the generalized Green--Kubo relation underlying Eq.~\eqref{eq:LRT_FDT_field}. This derivation is included to clarify the formal connection between the response-kernel formulation used in the Main Text and standard linear response theory. It is not required for the numerical extraction of the kernels, which is described in Materials and Methods, but provides the theoretical basis for identifying the measured equilibrium current correlations with field-level response functions.

The starting point is the assumption that the weakly perturbed system can be described by an effective time-dependent Hamiltonian~\cite{TOSL,zwanzig2001nonequilibrium},
\begin{equation}
\mathcal{H}^{\mathrm{neq}}(\vec{\Gamma},t)
= \mathcal{H}^{\mathrm{eq}}(\vec{\Gamma}) + \Delta\mathcal{H}(\vec{\Gamma},t),
\end{equation}
where \(\mathcal{H}^{\mathrm{eq}}\) is the equilibrium Hamiltonian and \(\Delta\mathcal{H}\) is a small perturbation that couples to conserved densities,
\begin{equation}
\Delta\mathcal{H}(\vec{\Gamma},t)
= - \sum_b \int_V \phi_b(\vec{r},t)\, \delta\rho_b(\vec{\Gamma},\vec{r})\,\mathrm{d}^3\vec{r}.
\end{equation}
Here \(\phi_b\) is the thermodynamic potential conjugate to the conserved density \(\rho_b\). The associated microscopic flux density \(\vec{j}_b\) is defined through the corresponding continuity equation.

The phase-space density $f^{\mathrm{neq}}(\vec{\Gamma},t)$ evolves according to the Liouville equation
\begin{equation}
\frac{\partial f^{\mathrm{neq}}}{\partial t}
= -\left\{\mathcal{H}^{\mathrm{neq}}, f^{\mathrm{neq}}\right\},
\end{equation}
where $\{\cdot,\cdot\}$ denotes the Poisson bracket. Introducing the perturbation $\Delta f = f^{\mathrm{neq}} - f^{\mathrm{eq}}$, where $f^{\mathrm{eq}}$ is the canonical equilibrium distribution associated with $\mathcal{H}^{\mathrm{eq}}$, one obtains, to first order in $\Delta\mathcal{H}$ and $\Delta f$,
\begin{equation}
\frac{\partial \Delta f}{\partial t}
+ \mathrm{i}\mathcal{L}^{\mathrm{eq}} \Delta f
= \left\{\Delta\mathcal{H}, f^{\mathrm{eq}}\right\},
\end{equation}
with $\mathrm{i}\mathcal{L}^{\mathrm{eq}} A = -\{\mathcal{H}^{\mathrm{eq}},A\}$ the equilibrium Liouville operator. The formal solution is
\begin{equation}
\Delta f(\vec{\Gamma},t)
= \int_{-\infty}^t \mathrm{e}^{-\mathrm{i}(t-s)\mathcal{L}^{\mathrm{eq}}}
\left\{\Delta\mathcal{H}(\vec{\Gamma},s), f^{\mathrm{eq}}(\vec{\Gamma})\right\}\,\mathrm{d}s.
\end{equation}
Using the canonical form of $f^{\mathrm{eq}}$ and standard manipulations, one finds
\begin{equation}
\left\{\Delta\mathcal{H}, f^{\mathrm{eq}}\right\}
= -\beta \sum_b \int_V \vec{j}_b(\vec{\Gamma},\vec{r})\cdot
\vec{\nabla}\phi_b(\vec{r},t)\,f^{\mathrm{eq}}(\vec{\Gamma})\,\mathrm{d}^3\vec{r},
\end{equation}
where $\vec{j}_b$ are the microscopic flux densities associated with the conserved densities $\rho_b$. Substituting this into the expression for $\Delta f$ and inserting into the nonequilibrium expectation value of a flux $\vec{j}_a$, one obtains
\begin{align}
\big\langle \vec{j}_a(\vec{r},t)\big\rangle_{\mathrm{neq}}
&= \int \vec{j}_a(\vec{\Gamma},\vec{r})\,\Delta f(\vec{\Gamma},t)\,\mathrm{d}\vec{\Gamma} \\
&= -\beta \sum_b \int_V \int_{-\infty}^t
\left\langle \vec{j}_a(\vec{r},t)\,
\vec{j}_b(\vec{r}',t') \right\rangle_{\mathrm{eq}}
\cdot\vec{\nabla}\phi_b(\vec{r}',t')
\,\mathrm{d}\vec{r}'\,\mathrm{d}t',
\end{align}
which is the desired linear response relation. Identifying the kernel in Eq.~\eqref{eq:LRT_field} yields the generalized Green--Kubo relation Eq.~\eqref{eq:LRT_FDT_field}.


\subsection*{Supplementary Text}\label{sup_text}

\subsubsection*{Thermodynamic force--flux conjugacy}
\label{sup_text:THERMO}

The derivation above establishes the dynamical part of the response theory: once conserved densities and their conjugate perturbing fields are specified, Hamiltonian linear response theory relates the corresponding flux response to equilibrium current correlations. It does not, by itself, determine which thermodynamic variables provide the most useful coarse-grained description of a confined fluid. This distinction is important at the nanoscale, where interfaces make the definition of fluxes, excess quantities, thermodynamic forces, and boundary conditions nontrivial.

The response-kernel formulation is therefore not tied to a unique choice of transported variables~\cite{yoshida2014generic}. Different flux bases may be introduced, provided that the conjugate thermodynamic forces are transformed consistently so that the entropy-production pairing is preserved. If the fluxes are transformed according to
\begin{equation}
    j'_a = \Gamma_{ab} j_b ,
\end{equation}
with summation over repeated indices, then the conjugate force gradients must transform contragrediently,
\begin{equation}
    \nabla \phi'_a
    =
    \left(\Gamma^{-T}\right)_{ab}
    \nabla \phi_b ,
\end{equation}
so that the bilinear force--flux pairing is unchanged,
\begin{equation}
    \sum_a j'_a \cdot \nabla \phi'_a
    =
    \sum_a j_a \cdot \nabla \phi_a .
\end{equation}
With this convention, the response kernel transforms covariantly as
\begin{equation}
    R'_{ab}(z,z',t)
    =
    \Gamma_{ac}\,
    R_{cd}(z,z',t)\,
    \Gamma_{bd}.
    \label{eq:R_basis_transform}
\end{equation}
The same transformation applies to the projected kernels \(K_{ab}\), \(G_{ab}\), \(M_{ab}\), and to the fully integrated Onsager matrix \(L_{ab}\). Here we restrict ourselves to spatially uniform changes of flux basis, such as the excess-current transformation used below. More general position-dependent transformations \(\Gamma(z)\) would introduce additional gradient and boundary terms in the force--flux pairing and are therefore not considered here.

This point is particularly relevant for confined fluids. Reservoir-imposed pressure, chemical-potential, temperature, and electric-potential differences do not necessarily correspond to independent local microscopic forces throughout the pore. Instead, they generate coupled perturbations whose projection onto local thermodynamic fields and effective boundary conditions can be representation-dependent. The role of the response-kernel framework is to keep the microscopic current correlations explicit while making these projections transparent.

In this sense, the nanoscale difficulty is not that linear response theory fails, but that the thermodynamic representation entering it becomes nontrivial. The response kernel provides a common microscopic object from which different coarse-grained descriptions may be constructed and tested. The problem of nanoscale irreversible transport is therefore shifted from asking whether Onsager symmetry holds for any particular projected representation to asking which thermodynamic representations faithfully preserve the underlying Hamiltonian symmetry structure under projection and coarse graining.

\subsubsection*{Excess flux basis}\label{sup_text:EXCESS}

In the Main Text, the full \(4\times4\) response kernel revealed a block structure among the particle, solute, heat, and charge transport channels. Here we show how this structure is modified when the solute and heat currents are expressed in an excess basis, obtained by subtracting the bulk advective contribution associated with the particle current.

We define the excess solute and heat fluxes as
\begin{equation}
j_s^{\mathrm{exc}}(z,t)
=
j_s(z,t)
-
\frac{\rho_s^{\mathrm{bulk}}}{\rho_n^{\mathrm{bulk}}}\,j_n(z,t),
\qquad
j_h^{\mathrm{exc}}(z,t)
=
j_h(z,t)
-
\frac{\rho_h^{\mathrm{bulk}}}{\rho_n^{\mathrm{bulk}}}\,j_n(z,t),
\end{equation}
while \(j_n\) and \(j_c\) are left unchanged. This transformation removes the contribution that would arise from the uniform advection of bulk solute and enthalpy by the particle flux, and therefore emphasizes deviations associated with interfacial structure and coupling.

Introducing the flux vector
\[
\mathbf{j}
=
\begin{bmatrix}
j_n \\ j_s \\ j_h \\ j_c
\end{bmatrix},
\qquad
\mathbf{j}^{\mathrm{exc}}
=
\begin{bmatrix}
j_n \\ j_s^{\mathrm{exc}} \\ j_h^{\mathrm{exc}} \\ j_c
\end{bmatrix},
\]
the transformation can be written in matrix form as
\begin{equation}
\mathbf{j}^{\mathrm{exc}} = \Gamma\,\mathbf{j},
\qquad
\Gamma =
\begin{bmatrix}
1 & 0 & 0 & 0 \\
-\rho_s^{\mathrm{bulk}}/\rho_n^{\mathrm{bulk}} & 1 & 0 & 0 \\
-\rho_h^{\mathrm{bulk}}/\rho_n^{\mathrm{bulk}} & 0 & 1 & 0 \\
0 & 0 & 0 & 1
\end{bmatrix}.
\end{equation}

This excess representation is not unique: it is one possible linear thermodynamic change of basis, chosen here because it subtracts the bulk advective contribution of the particle current from the solute and heat currents. To preserve the bilinear flux--force structure, the conjugate thermodynamic forces transform according to
\begin{equation}
\nabla\phi^{\,\mathrm{exc}}
=
\Gamma^{-T}\nabla\phi,
\end{equation}
so that the response kernels transform as
\begin{equation}
\mathbf{R}^{\mathrm{exc}} = \Gamma\, \mathbf{R}\, \Gamma^T,
\end{equation}
and similarly for the projected kernels \(\mathbf{K}\), \(\mathbf{G}\), \(\mathbf{M}\), and the integrated Onsager matrix \(\mathbf{L}\).

\begin{figure}
    \centering
    \begin{minipage}[c]{0.60\textwidth}
        \centering
        {\Large\bfseries A}\\[0.1em]
        \includegraphics[width=\textwidth]{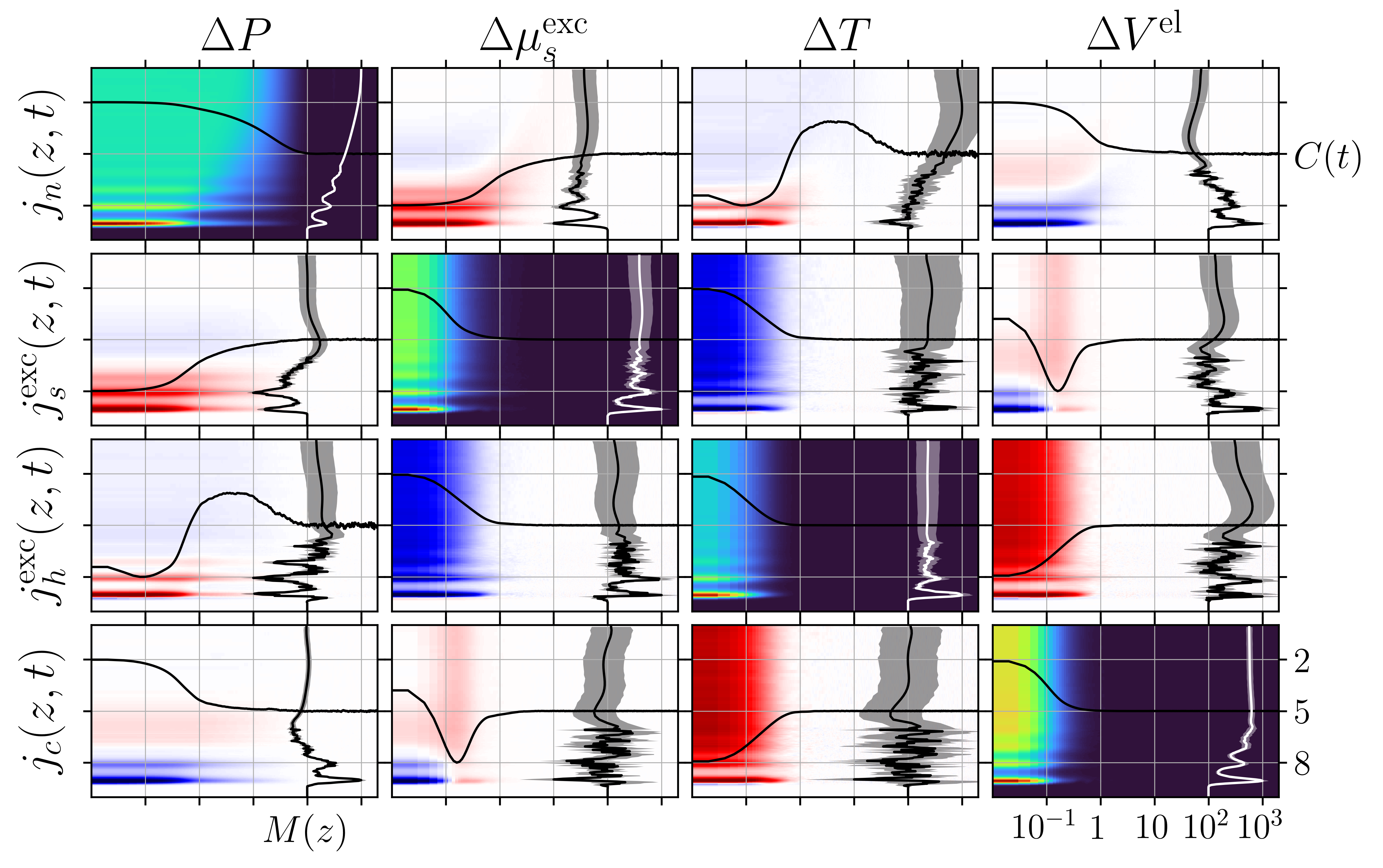}
    \end{minipage}
    \hfill
    \begin{minipage}[c]{0.38\textwidth}
        \centering
        {\Large\bfseries B}\\[0.1em]
        \includegraphics[width=1.\textwidth]{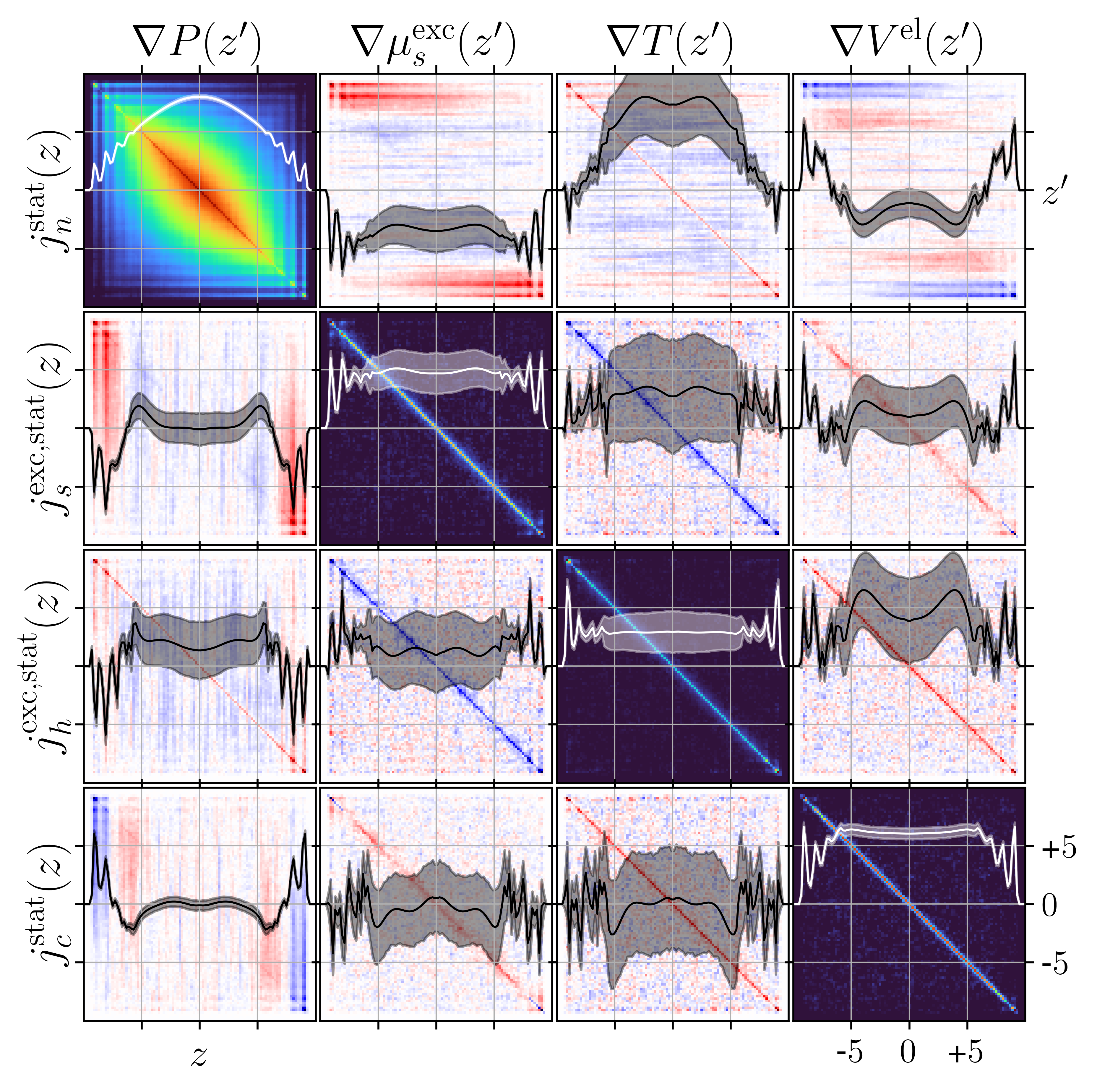}
    \end{minipage}
    \caption{\begin{small}
    \textbf{Response kernels in the excess solute and heat flux basis.}
    Same layout as in Fig.~\ref{fig:full_matrix}: (\textbf{A}) local--global space--time kernels \(K_{ab}(z,t)\), with the corresponding stationary profiles \(M_{ab}(z)\) and global temporal projections \(C_{ab}(t)\); (\textbf{B}) stationary local--local response functions \(G_{ab}(z,z')\). The transformed fluxes are \(j_s^{\mathrm{exc}} = j_s - (\rho_s^{\mathrm{bulk}}/\rho_n^{\mathrm{bulk}})j_n\) and \(j_h^{\mathrm{exc}} = j_h - (\rho_h^{\mathrm{bulk}}/\rho_n^{\mathrm{bulk}})j_n\), which subtract the bulk advective contributions to the solute and heat currents. Diagonal components are shown with a positive colormap, whereas off-diagonal components use a symmetric diverging colormap. Removing the advective projection suppresses the broad hydrodynamic component of the solute and heat responses and reveals faster, more localized residual contributions, particularly in the \(ss\) and \(hh\) sectors. The qualitative block structure of the response matrix is preserved, but the amplitudes and visual prominence of individual entries depend on the chosen flux basis. For visualization only, the profiles were interpolated across the pore center in panels where symmetry and low signal-to-noise made the central values difficult to display.
    \end{small}}
    \label{fig:sup_excess}
\end{figure}

Figure~\ref{fig:sup_excess} displays the resulting response kernels in this excess basis. Compared with the original representation of Fig.~\ref{fig:full_matrix}, the transformation strongly suppresses the long-lived hydrodynamic component in the solute and heat channels. This is most visible in the diagonal \(ss\) and \(hh\) entries: in the bare basis, these responses contain a broad advective contribution inherited from the induced particle current, whereas in the excess basis the remaining correlations are shorter lived and more spatially localized. The corresponding stationary local--local responses \(G_{ss}^{\mathrm{exc}}(z,z')\) and \(G_{hh}^{\mathrm{exc}}(z,z')\) are therefore much closer to diagonal than in the bare representation, indicating that the residual solute and heat responses are dominated by local microscopic, ionic, or thermal relaxation rather than by pore-scale momentum diffusion.

This comparison illustrates the representation-dependence of projected transport kernels. The excess basis should not be interpreted as uniquely more fundamental than the bare flux basis; it is a diagnostic projection that removes one chosen bulk advective contribution. The robust feature is the separation between hydrodynamic, ionic, and mixed interfacial pathways. The detailed appearance of individual projected kernels, however, depends on the flux basis and on the corresponding conjugate thermodynamic forces.

\subsubsection*{Limitations of the present approach}\label{sup_text:LIM}

Several limitations of the present approach should be emphasized. First, the response matrix is defined only after choosing a set of microscopic observables and their conjugate thermodynamic perturbations. It therefore depends on the flux--force representation, on the definition of excess currents, and on the projections used to reduce the full nonlocal kernel. This issue is particularly important for heat transport, where microscopic energy and heat currents may contain convective, virial, and interaction-energy contributions. The heat-related kernels should therefore be interpreted as response functions within a specified microscopic representation, rather than as a unique decomposition of all possible thermal transport mechanisms.

Second, the charged Lennard--Jones nanopore studied here is a deliberately minimal model. It isolates the structure of the response-kernel formalism, but does not include molecular water, orientational polarization, hydrogen bonding, ion-specific hydration, surface chemistry, charge regulation, or dielectric heterogeneity. In realistic aqueous electrolytes, additional molecular modes such as water rotational relaxation or hydration-shell rearrangements may be spectrally separated from ionic diffusion and hydrodynamic relaxation. Their coupling to mass, salt, charge, or heat transport may therefore be weaker, delayed, or more structured than in the present model. The results should thus be read as a microscopic proof of principle for reconstructing space--time-resolved coupled transport, rather than as quantitative predictions for a specific material interface.

Third, the present implementation is statistically and computationally demanding. We deliberately reconstructed the kernels by direct, essentially brute-force sampling of equilibrium flux correlations. This choice keeps the analysis transparent and avoids imposing a preconceived structure on the response, but it is not statistically optimal. High-dimensional correlation functions involve a bias--variance tradeoff: direct estimates have low modeling bias but high variance, especially for off-diagonal channels, interfacial features, and long-time tails. Moreover, space--time correlations require continuous trajectories long enough to resolve the relevant relaxation times, and the full nonlocal kernel scales rapidly with the number of observables, spatial bins, and time lags. Future work should therefore combine physical constraints with improved inference methods, including symmetry-constrained reconstruction, reduced bases, Bayesian regularization, Gaussian-process models, neural operators, or physics-informed regression. Such approaches could transform response kernels from diagnostic visualizations into quantitative constitutive objects with controlled uncertainty.

Finally, the formalism used here is restricted to linear response around equilibrium in a closed system. The kernels describe the first-order response to weak thermodynamic driving and cannot, by themselves, capture strongly nonlinear phenomena such as field-induced structural changes, concentration polarization, nonlinear electro-osmosis, Joule heating, adsorption hysteresis, chemical reactions, or ion depletion. Furthermore, many experimentally relevant nanofluidic and electrochemical systems are open: they exchange particles, heat, and charge with reservoirs and sustain finite gradients in nonequilibrium steady states. The equilibrium kernels obtained here can serve as local microscopic building blocks for such descriptions, but additional assumptions are required to embed them into open-system boundary-value problems or nonlinear constitutive laws.



\end{document}